\documentclass[aps,prb,reprint,superscriptaddress,showpacs,twocolumn]{revtex4-2}
\usepackage{color}
\usepackage{graphicx}
\usepackage[american]{babel}
\usepackage[T1]{fontenc}
\usepackage{amsmath}
\usepackage{amssymb}
\usepackage{amstext}
\usepackage{amsthm}
\usepackage{latexsym}
\usepackage{verbatim}
\usepackage{natbib}
\usepackage{array}
\usepackage{color} 

\usepackage{ulem}   

\def\CoSnS{Co$_{3}$Sn$_{2}$S$_{2}$}
\def\Bbr{$B_{br}$}
\def\BLT{$B_{LT}$}
\def\Bint{$\vec{B}_{int}$}
\def\deg{$^{\circ}$}
\usepackage[draft]{todonotes}   

\usepackage{xspace}

\begin{document}

\title {NMR study of the local magnetic order in the kagome Weyl semimetal Co$_3$Sn$_2$S$_2$ }

\author{Irek Mukhamedshin}
\author{Pawel Wzietek}
\author{Fabrice Bert}
\author{Philippe Mendels}
\affiliation {Universit\'{e} Paris-Saclay, CNRS, Laboratoire de Physique des Solides, 91405, Orsay, France.}

\author{Anne Forget}
\author{Doroth\'{e}e Colson}

\affiliation{Service de Physique de l'Etat Condens\'{e}, Orme des Merisiers, CEA Saclay, CNRS-URA 2464, 91191 Gif sur Yvette Cedex, France}

\author{V\'{e}ronique Brouet}
\affiliation {Universit\'{e} Paris-Saclay, CNRS, Laboratoire de Physique des Solides, 91405, Orsay, France.}

\begin{abstract}

A magnetic Weyl semimetal presents the intriguing possibility of controlling topological properties through magnetic order. The kagome compound \CoSnS~has emerged as one of the most thoroughly characterized magnetic Weyl semimetals, yet the potential coexistence of a ferromagnetic state below $T_c$ = 172~K with a non-collinear antiferromagnetic phase or a glassy state remains unresolved. We employ $^{59}$Co NMR to gain a local perspective on the magnetic order. The magnetic and electric field gradient tensors at room temperature are determined by fitting the NMR spectra using evolutionary algorithms. Zero-field NMR measurements reveal that all Co sites are equivalent in the magnetic phase at low temperatures and up to 90~K. The local magnetic field follows in intensity the macroscopic magnetization as a function of temperature and is tilted from the c-axis by a few degrees toward the nearest triangle center. Above 90~K, a shoulder appears on the low-field side, which we attribute to a preferential tilting of the local field in one direction, breaking the equivalence between the three Co sites of the kagome structure. We rule out any coexistence with an in-plane antiferromagnetic phase and suggest instead that in-plane ferromagnetic-like moments appear above 90~K and play an increasing role in the magnetic order up to the magnetic transition.

 \end{abstract}

\date{\today}

\maketitle

Combining magnetism and topology is an interesting perspective of today's condensed matter research. As a fictitious analog of a magnetic field, the Berry curvature found in topological materials may give new ways to manipulate electrons \cite{Nakatsuji2022} and stabilize exotic states of matter such as the long sought quantum anomalous Hall effect\cite{HaldanePRL88}. In the past decade, it was realized that kagome metals based on Mn, Fe or Co offer a very good platform for such an investigation. The 3d transition metals are prone to magnetism and the kagome lattice intrinsically has a topological band structure \cite{GuoPRB09}. An original example might be Na$_{2/3}$CoO$_2$, where the Na order differentiates a Co kagome sublattice in triangular CoO$_2$ layers \cite{GilmutdinovPRB21}. More recently, a wealth of interesting phases have been discovered in intermetallic stannides \cite{YinHasan22}. For example, Mn$_3$Sn displays an anomalous Hall effect at room temperature \cite{NakatsujiNature2015}, which is unconventional as it is not due to ferromagnetism, but to the Berry curvature associated with the  120$^{\circ}$ non-collinear AFM order on each triangle of the kagome structure. In \CoSnS, a magnetic Weyl semimetal is formed, displaying a chiral anomaly\cite{LiuNatPhys18} and Fermi arcs \cite{LiuScience19}.

\CoSnS~crystallizes in a rhombohedral structure (space group, R$\bar{3}$m), with Co$_3$Sn kagome planes. It is ferromagnetic (FM) below $T_c=172~K$, with magnetic moments $M\sim0.3\mu_B$ per Co, preferentially aligned along the c-axis perpendicular to the kagome planes \cite{SchnellePRB13}. Such a value is well explained by the small filling of the semimetallic fully spin-polarized bands \cite{LiuNatPhys18}. Two bands of opposite symmetry form a nodal ring in the mirror symmetry plane, which is gapped by spin-orbit coupling, except at two Weyl points. The very large anomalous Hall angle \cite{LiuNatPhys18} is believed to be enhanced by the proximity of the Weyl points to the Fermi level \cite{LiuScience19,LohaniPRB23}. 

Manipulating the topological phase via magnetism is an appealing possibility. Although the FM state dominating in \CoSnS~apparently releases the frustration inherent to the kagome lattice~\cite{Mendels16} and suggests a simple magnetic behavior, it was predicted theoretically that the decrease of the ordered magnetic moment, as expected near $T_c$ or when doped, could destabilize the FM state and lead to the emergence of spin-spiral and AFM orders \cite{SolovyevPRB22}. Interestingly, a muon spin rotation ($\mu$SR) study \cite{GuguchiaNatCom20} detected magnetic oscillations with two frequencies above 90~K, suggesting that the magnetic state is indeed evolving and is more complicated than previously anticipated. The second frequency was attributed to an AFM non-collinear phase coexisting with the FM state and reaching a fraction of 80\% just below the transition. The direct detection of the AFM state by neutron scattering is difficult because it is a q=0 order. There are 2 different AFM orders compatible with the space group of \CoSnS, with spins pointing to a neighboring hexagon ($\Gamma_1^+$) or a neighboring triangle ($\Gamma_2^+$) \cite{SohBoothroydPRB22}. The first one, originally proposed by the $\mu$SR study, has a different symmetry with respect to FM order and could be ruled out by neutron experiments \cite{SohBoothroydPRB22}. The "umbrella tilting" one can be obtained within the same symmetry by continuous tilting of the spins away from the $c$ axis by an angle $\theta$ towards the triangle center. It remains compatible with neutron data \cite{SohBoothroydPRB22} and a tilt angle strongly dependent on temperature and on In doping was reported \cite{NeubauerNPJQM22}. This spin configurations is a chiral object that was proposed to give rise to topological Hall effect in In-doped \CoSnS \cite{KassemPhD16}. Additionally, anomalies were reported near 130~K in the susceptibility and domain wall dynamics \cite{KassemPRB17,SugawaraPRM19,LeeNatCom22} and it was argued that a small net in-plane magnetic component appears below $T_P$=130~K \cite{ZivkovicPRB22}. 

\vskip 0.5cm

In order to unveil the still elusive nature of this competing magnetic order, we report here a $^{59}$Co NMR study of \CoSnS~both in paramagnetic (PM) and FM states. Generally, a nuclear spin $\vec{I}$ ($I$>1/2), couples to its magnetic  and anisotropic charge environment\cite{Abragam} through  the Hamiltonian $\mathcal{H}$.
\begin{equation}
	\mathcal{H}=-\gamma \hbar \vec{I}\cdot\vec{B}_{int}+\frac{eQ}{2I(I-1)}\vec{I}\hat{V}\vec{I},  \label{eq:Hamiltonian}
\end{equation}%
where $\gamma $ is the gyromagnetic ratio and Q the quadrupole moment. The internal magnetic field in the PM phase is $\vec{B}_{int}=(1+\hat{K})\vec{B}_{0}$, with $\vec{B}_0$ the applied magnetic field and  $\hat{K}$ the magnetic shift tensor. A strong electric field gradient (EFG) typically splits the NMR spectrum into a central line and (2I-1) quadrupole satellites \cite{Abragam}, i.e. 7 lines for $^{59}$Co ($I$=7/2). In magnetic phases, the existence of strong internal fields at the nuclear site enables to study the magnetic order without any perturbation from an applied magnetic field (zero field NMR, ZF-NMR). The signal is essentially sensitive to the on-site electronic moment M, through \Bint=$\hat{A}\cdot\vec{M}$, where the hyperfine tensor $\hat{A}$ describes the coupling between the nuclear spin and electrons \cite{sup}. It is then more direct than $\mu$SR that averages the dipolar fields of many Co sites. Also contrary to the neutron experiments NMR is not hampered by a q=0 long range magnetic order. Moreover, the NMR signal intensity is usually simply proportional to the number of resonating Co nuclei, minimizing the impact of defects or sample boundaries. In the FM state, it can be complicated by a large enhancement factor, different for bulk and domain walls \cite{TurovPetrov}. However, we found that the enhancement factor is near 1 \cite{sup} in \CoSnS, due to the large magnetic anisotropy. Therefore, the signal from the narrow domain walls \cite{SugawaraPRM19} is safely negligible, which greatly simplifies the analysis compared to macroscopic magnetization measurements. 

\vskip 0.5cm

\begin{figure}[tbp]
\center
\includegraphics[width=0.9\linewidth]{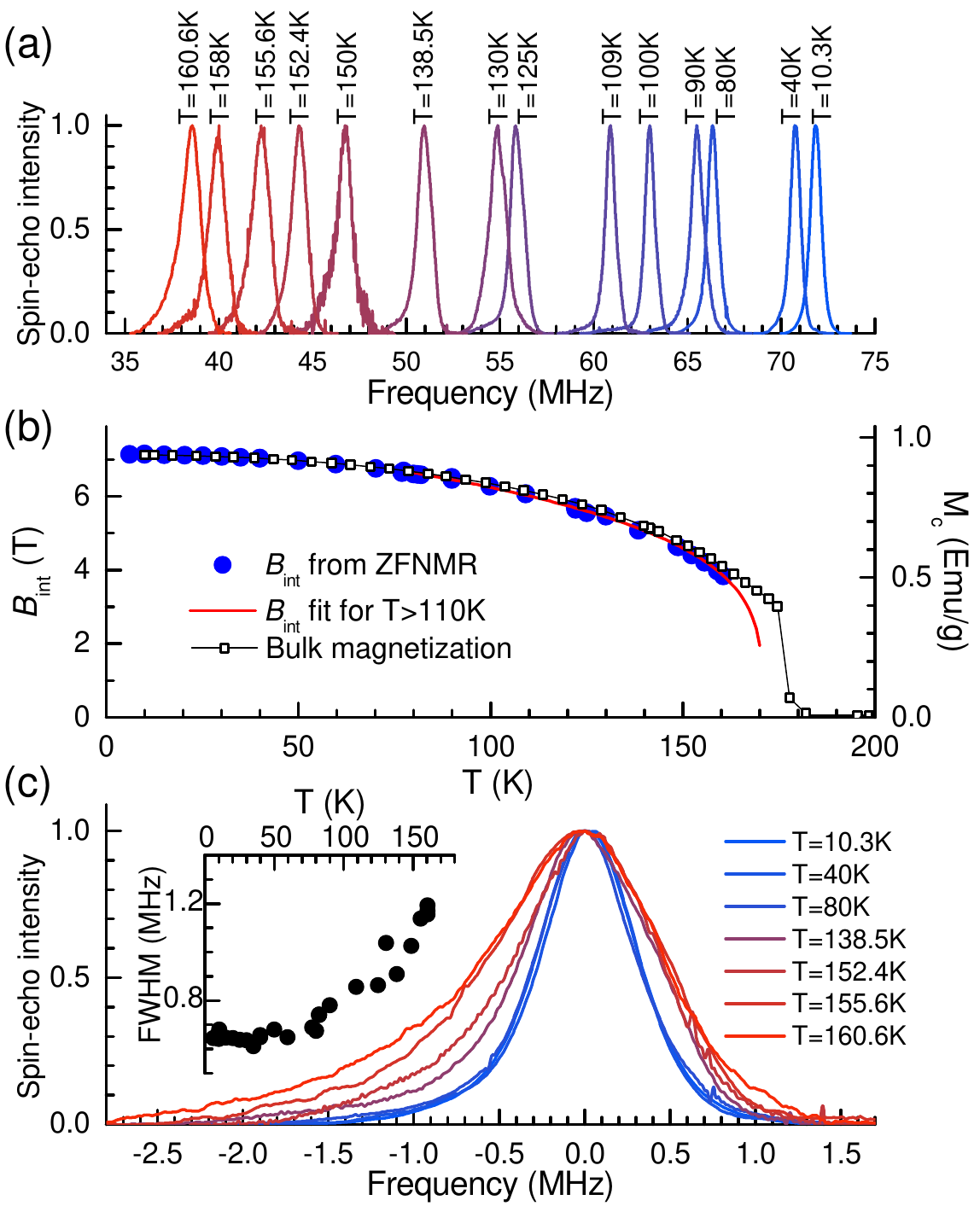}
\caption{(a) $^{59}$Co ZF-NMR spectra in the magnetically ordered state of \CoSnS~(powder sample), normalized to their maximum. (b) Temperature dependence of the internal magnetic field $B_{int}$ (blue filled circles, left axis) deduced from the position of the spectra maximum (we used $\gamma /2\pi =$ 10.054~MHz/T). The red line is a fit above 110~K to $B_{int}(T)=7.8(1-(T/171))^{0.25}$. Right axis (open squares) : SQUID magnetization measured on a single-crystal in a field $B=2.5$~mT (B$\parallel$c, field cooled). (c) Set of $^{59}$Co ZF-NMR normalized spectra and shifted to the zero frequency. Inset: temperature dependence of the FWHM.}
\label{fig:FigZF}
\end{figure}

Single crystals of \CoSnS~were grown by the self-flux method \cite{LohaniPRB23} and ground into powders for ZF-NMR experiments, where only the local orientation of the electronic moments matters \cite{sup}. Figure~\ref{fig:FigZF}(a) displays $^{59}$Co ZF-NMR spectra taken from 160~K, just below the magnetic transition, down to 10~K. At each temperature, we observe a single and relatively narrow line strikingly much simpler than the 7 lines expected from quadrupolar effects. The position of the line maximum is strongly temperature dependent. It is reported on the left axis of Fig.~\ref{fig:FigZF}(b) and reflects the development of the magnetization in the sample, similarly to the bulk magnetization (right axis). There is a small deviation between the two curves close to  $T_c$, which can be attributed to the influence of the external magnetic field on the bulk magnetization. 

In order to detect a more subtle evolution of the spectra, we superimpose in Fig.~\ref{fig:FigZF}(c) some of $^{59}$Co ZF-NMR spectra by shifting them to zero frequency and normalizing them to their maximum. The lineshape does not change for $T<90$~K, but a shoulder develops on the low frequency side at higher temperatures. The inset of Fig.~\ref{fig:FigZF}(c) shows that the full width at half maximum (FWHM), constant below 90~K, starts increasing above this temperature. This change occurs on a very similar temperature window as the anomalies in the magnetic behavior of \CoSnS~reported previously \cite{GuguchiaNatCom20,ZivkovicPRB22}, so that it is natural to wonder how these observations are related and what could be their origin.

\vskip 0.5cm

\begin{figure}[tbp]
	\center
	\includegraphics[width=0.9\linewidth]{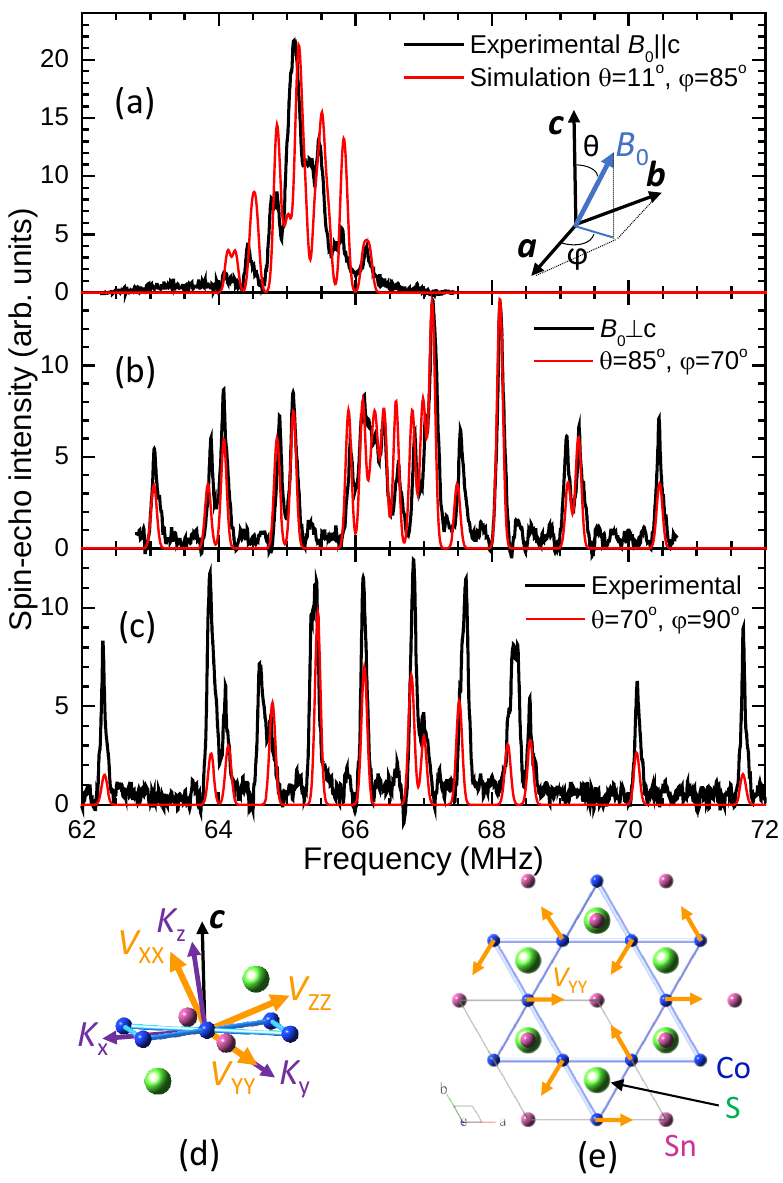}
	\caption{(a-c) $^{59}$Co NMR spectra (black lines) taken in the PM state of a Co$_3$Sn$_2$S$_2$ single crystal at $T=300~K$ for the indicated orientions of the external magnetic field $\vec{B}_{0}$. Panel (c) shows the broadest spectrum we could obtain, which corresponds then to the situation $B_{0} \parallel V_{ZZ}$ for one Co site. The red lines are results of simulations described in the text. (d) Sketch of the local structure around one Co site with the main axis of magnetic shift tensor $\hat{K}$ (purple) and EFG tensor $\hat{V}$ (orange) determined by our simulation. (e) 2D view of the kagome plane. S and Sn atoms are alternatively above and below the center of the triangles [see also panel (d)]. The in-plane principal component $V_{YY}$ of the EFG tensor is sketched at the 3 inequivalent Co sites.}
	\label{fig:Fig300K}
\end{figure}

To clarify the origin of the ZF-NMR evolution, it is necessary to get a better understanding of the relative contribution of magnetic and quadrupolar tensors to the lineshape (see Eq.~\ref{eq:Hamiltonian}). We determine them in the PM phase by measuring $^{59}$Co NMR spectra (black lines in Fig.~\ref{fig:Fig300K}) at 300~K for a single crystal and different orientations of the applied magnetic field $\vec{B}_0$~\cite{sup}. The spectra consist of many relatively narrow and well resolved lines, the number and position of which strongly depend on the orientation of $\vec{B}_0$. The lines are distributed on a much wider frequency window when $\vec{B}_0$ is closer to the kagome plane ($\theta$=$90$\deg), where up to 21 lines can be observed.  

There are 3 inequivalent Co sites in a kagome plane, for which the local frames are rotated around $c$ by 120$^{\circ}$ from each other, as sketched in Fig.~\ref{fig:Fig300K}(e). Since 7 lines are expected from quadrupolar interactions for a nuclear spin $I$=7/2, this explains that up to 21 lines can be observed. In perturbation theory \cite{Abragam}, the distance between outer satellites in the $^{59}$Co NMR spectra depends on the orientation of the internal magnetic field \Bint~with respect to the principal axis of EFG tensor, described by the spherical angular coordinates $\theta_V$ and $\varphi_V$.
\begin{equation}
	\Delta \nu =3\nu _{Q}(3\cos ^{2}\theta_V -1+\eta \sin ^{2}\theta_V
	\cos 2\varphi_V)  \label{eq:SatelDist}
\end{equation}
where the quadrupolar frequency  $\nu_{Q}$ is a measure of the largest EFG component $V_{ZZ}$ and the anisotropy parameter $\eta =(V_{XX}-V_{YY})/V_{ZZ}$ decribes the in plane components of the EFG. On one hand, the maximum splitting is expected in the direction of $V_{ZZ}$. As it is observed around $\theta \approx 70 ^{\circ}$, $V_{ZZ}$ is probably oriented close to this direction and not along the c-axis as found in other Co layered materials such as Na$_x$CoO$_2$ \cite{H67_CoNMR,PhysicaB2015} or CoSn \cite{HuangPRL22}. Indeed the ions distribution around one Co, sketched in Fig.~\ref{fig:Fig300K}(d), is very asymmetric with S atoms alternatively above and below the Co triangles probably roughly shaping the direction of $V_{ZZ}$. On the other hand, there is a "magic angle", for which the angular dependence is minimal and all satellites collapse to the same frequency. We can anticipate that the spectrum in Fig.~\ref{fig:Fig300K}(a), with $\vec{B}_0$ nearly parallel to $c$, is close to this condition.

To solve the complex problem of obtaining magnetic and EFG tensors from these spectra, we acquired a set of 12 $^{59}$Co NMR spectra for different crystal orientations and fitted the positions of the lines in all 12 spectra simultaneously. A numerical diagonalization of the full Hamiltonian (Eq.~\ref{eq:Hamiltonian}) was used to obtain the resonance frequencies.  A multi-parameter fit was carried using a genetic algorithm called Differential Evolution \cite{storn_price}. This method probes large areas of configuration space and is not subject to fall into local minima in contrast to conventional gradient-based methods. It is therefore very effective for global minimization problems involving many parameters \cite{wormington,das2010de}. In order to respect the symmetry constraints, we fix the $K_y$ and $V_{YY}$ directions at one Co site along the $C_2$ axis of rotation in the structure \cite{XuFelserPRB18} and let the 2 other components vary in the perpendicular mirror symmetry plane [see Fig.~\ref{fig:Fig300K}(d-e)]. To describe the two other inequivalent Co sites of the kagome sublattice, we rotate the tensors by 120$^{\circ}$ and 240$^{\circ}$ around $c$. 

We obtain a satisfactory simulation of all experimental spectra, as shown by red lines in Fig.~\ref{fig:Fig300K}(a-c), with the following parameters:
\begin{eqnarray}
	&&K_x = -0.06\%; K_y = -1.7\%; K_z = -3\%; \nonumber \\
	&&\nu_Q = 1.56~\text{MHz}; \eta = 0.36; \nonumber \\
	&&\theta_{K_0} = 355 ^{\circ}; \theta_{V_0} = 67 ^{\circ}; \nonumber
\end{eqnarray}	

The principal axis of the two tensors, given by Euler angles $\theta_{V_0}$ and $\theta_{K_0}$, do not coincide, as represented in Fig.~\ref{fig:Fig300K}(d). In order to reproduce the linewidths, we used a convolution by a Gaussian function with FWHM~=~0.1~MHz. The discrepancy between the relative intensities of the satellites and the central lines can be explained by the fact that the amplitude of RF pulses is not optimized for each transition \cite{ManPRB}. 

\vskip 0.5cm

\begin{figure}[tbp]
	\center
	\includegraphics[width=0.8\linewidth]{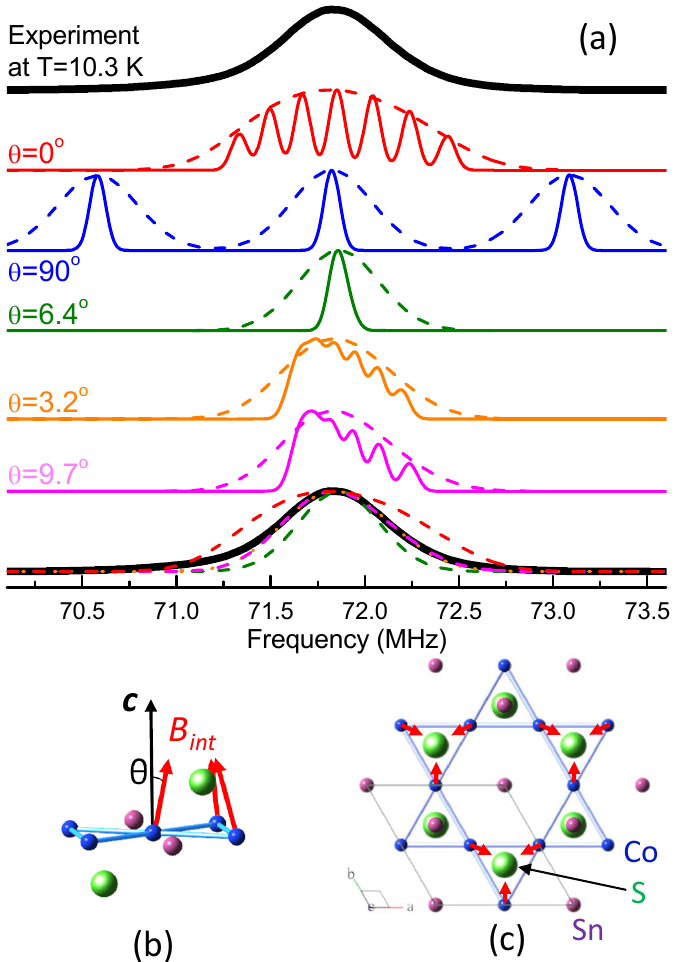}
	\caption{(a) Black thick line: $^{59}$Co ZF-NMR spectra at $T=10~K$. Color lines : computer simulations for different values of the $\theta$ angle between \Bint~and $c$-axis [see panel (b)]. For solid color lines, a broadening of 0.1~MHz is used (as for Fig.~\ref{fig:Fig300K}), whereas for dashed color lines, it is 0.5~MHz. At the bottom, the experimental spectrum is superimposed with some of the simulations. (b) "Umbrella" structure of \Bint~at the 3 Co sites. (c) Schematic representation of the projection of \Bint~(red arrows) onto the kagome plane.}
	\label{fig:Fig10K}
\end{figure}

We now reexamine the lineshape of the $^{59}$Co ZF-NMR spectra at $T<90~K$. As the structure does not change significantly at the transition \cite{ZhangJACS22}, we expect that the orientation and symmetry of the EFG tensor will not change, even though $\nu_Q$ could slightly evolve \cite{H67_CoNMR,CuS_PRB09}. The absence of quadrupole satellites characteristic of the low temperature spectra suggests that \Bint~is oriented close to the magic angle with respect to V$_{ZZ}$, as when $\vec{B}_0$ is oriented near $c$ in the PM case. In fact, we found that the spectrum width is so sensitive to the orientation of \Bint~that it can be used to get precise information about it. In Fig.~\ref{fig:Fig10K}, we show the experimental ZF-NMR spectrum at $T=10~K$ (thick black line, top and bottom), together with simulations for different orientations of \Bint. We restrict the orientation of \Bint~to the mirror symmetry plane, as other orientations would break the symmetry, which is needed in particular to form the Weyl points \cite{ZhangPRL21}. For $\theta = 0 ^{\circ}$, the simulation yields the classic picture with seven $^{59}$Co NMR lines (solid line with the same broadening as used in Fig.~\ref{fig:Fig300K}). Even if this structure is smoothed by an additional broadening (dashed line), the distance between the outer satellites is obviously too large to fit the width of the experimental spectrum. In the $\theta = 90 ^{\circ}$ case the distance between the satellites is much larger than the spectral width, even for the first transition (the other four satellites are out of the figure frequency range). This definitely rules out an in-plane AF order of \Bint. The $\theta = 6.4 ^{\circ}$ case gives the narrowest spectrum, as it corresponds to the "magic angle", but it is too narrow to fit the experiment. Spectra at $\theta = 3.2 ^{\circ}$ and $9.7 ^{\circ}$ give the best match, as shown at the bottom of the figure. 

The fact that only one set of parameters is enough to reproduce the experimental spectrum indicates that the 3 Co sites are equivalent. This implies that \Bint~has the same value and orientation with respect to the local frame for each Co, forming the umbrella structure described in introduction and sketched in Fig.~\ref{fig:Fig10K}(b). The distribution of the electronic moments $\vec{M}$ obeys the same symmetry constraints as \Bint, but in case of anisotropic $\hat{A}$, the tilting of $\vec{M}$ could be slightly larger \cite{sup}.

\vskip 0.5cm

\begin{figure}[tbp]
	\center
	\includegraphics[width=0.9\linewidth]{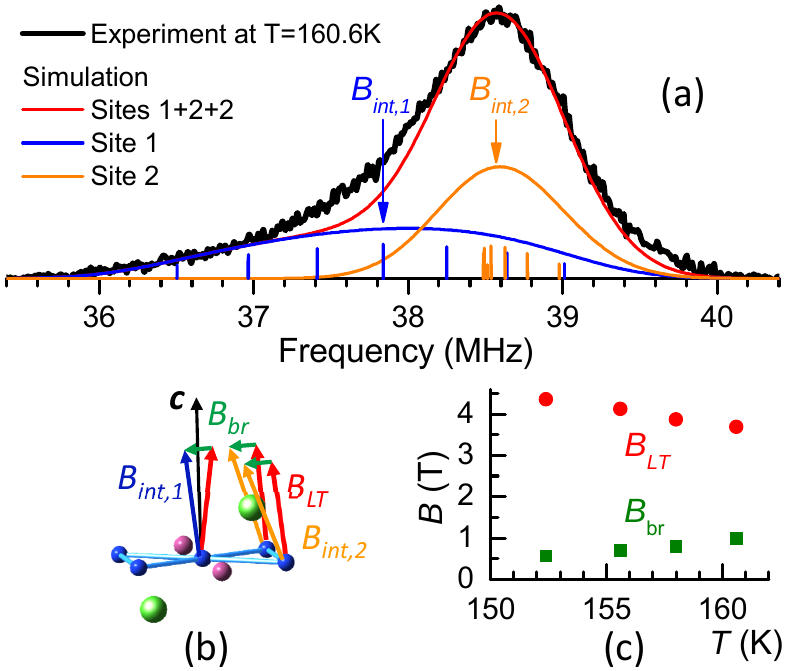}
	\caption{(a) Black thick line : $^{59}$Co ZF-NMR spectra at $T=160~$K. Simulations including a local symmetry breaking between site 1 and site 2 are shown by color thin lines (vertical bars indicate quadrupole satellites and the arrow the central value of $B_{int}$). (b) Schematic representation of the on-site magnetic fields for the three nonequivalent Co sites. The $B_{LT}$ field (red arrow) corresponds to the umbrella component, as found at low temperatures. The in-plane $\vec{B}_{br}$ field (green arrow) is the same for all Co sites, hence breaking the 3-fold rotation symmetry between the sites. The total field is shown as \Bint$_{,1}$~for site 1 (blue) and \Bint$_{,2}$~for the two sites 2 (magenta). (c) Temperature dependence of the $B_{LT}$ and $B_{br}$ fields, obtained in simulations. }
	\label{fig:Fig160K}
\end{figure}

Understanding this, we now turn to the evolution of the spectra above 90~K, namely the gradual appearance of a low-frequency shoulder highlighted in Fig.~\ref{fig:FigZF}(c). In Fig.~\ref{fig:Fig160K}(a), we show the $^{59}$Co ZF-NMR spectrum taken at $T=160~$K with the most prominent low frequency shoulder. This shoulder cannot be explained by an emergence of another phase with a completely different magnetic order, as any large change of the direction of \Bint~would lead to a drastic change in the lineshape, as demonstrated by the previous simulations. In particular, a coexistence with an AF phase, as suggested by $\mu$SR \cite{GuguchiaNatCom20}, would give a second distribution with satellites at very distinct positions, which are not observed. For the same reason, it cannot be associated to domain walls, as the magnetization would be strongly tilted or reduced in domain walls. Moreover, only a few Co sites would be affected, whereas the shoulder corresponds to a significant intensity. 

Varying the direction of \Bint~similarly for the 3 sites can produce some asymmetry and broadening (see Fig.~\ref{fig:Fig10K}), but it cannot produce a shoulder as it remains centered on roughly the same value. We have to break the symmetry between the 3 Co sites to create a shoulder. Indeed, we have found that the spectra can be decomposed into 2 lorentzians with a weight close to 1:2. 
One simple way to break the symmetry is to tilt the moment away from the symmetry plane for some of the Co sites. One solution depicted in Fig.~\ref{fig:Fig160K}(b) is to add to the umbrella component found at low temperature, which we now call $\vec{B}_{LT}$, an in-plane component $\vec{B}_{br}$ (green arrow) in one of the high-symmetry direction. This will reduce $\theta$ for one site (Site 1), but increase it for the two other sites (Site 2) and $B_{int}$ becomes larger for site 2. We note that a subtle symmetry breaking\cite{ZhangJACS22} was reported to appear below Tc, which does create inequivalent Co sites as detected here. However, the two are probably not related as the structural distortion increases at low temperatures, contrary to our magnetic symmetry breaking that disappears. In Fig. \ref{fig:Fig10K}(a), a spectrum simulation is presented at 10~K, using the same procedure as before, which reproduces quite well the experiment for \BLT=3.7~T ($\theta = 3.2 ^{\circ}$) and \Bbr=1~T ($\theta = 90 ^{\circ}$). We fitted all spectra where the shoulder is significant ($150<T<160$~K) and show in Fig.~\ref{fig:Fig160K}(d) the temperature dependence of the internal fields. $B_{br}$ gradually increases as the temperature rises towards $T_C$, while \BLT~decreases. 

As ZF-NMR is mainly sensitive to the on-site magnetic moment, we can assert that the value of this in-plane component reaches significant values close to $T_c$ (at least a quarter of that along $c$ \cite{sup}), but we cannot say how it is ordered on the long distance. ZF-NMR would be consistent not only with a uniform direction of $\vec{B}_{br}$ but also for example with a 120\deg~rotation of $\vec{B}_{br}$ from one triangle to the next, or at least domains with different orientations, so that it cannot be detected on the macroscopic scale.  Indeed, low-field magnetization measurements \cite{ZivkovicPRB22} only detect an in-plane magnetization component $M_p$ below $T_P$=128~K, precisely when our $B_{br}$ vanishes. There is no contradiction, as the very small $M_p$ value (several orders of magnitude less than along $c$) would not affect our measurement. The very existence of an in-plane component hints nonetheless, in our opinion, at the importance of this degree of freedom. Similarly, the complex behavior of hysteresis loops \cite{LachmanAnalytisNatCom20,MenilCondMat24} and the change of domain wall motion \cite{LeeNatCom22} in the same temperature range could be connected with the appearance of this in-plane order, although the exact relation is still unclear. Our NMR results are quite similar to the $\mu$SR ones as both detect two internal field values above 90~K. However the coexistence of two distinct magnetic phases originally proposed \cite{GuguchiaNatCom20} is inconsistent with our NMR data, which require a smooth evolution of the magnetic order. The differentiation of Co sites 1 and 2 we suggest likely breaks the equivalence of the various muon stopping sites, naturally leading to two distinct frequencies.

  
In summary, we have found that the ZF-NMR spectra are extremely sensitive to the direction of the internal magnetic field at Co sites \Bint, through the distribution of quadrupolar satellites. This is an unexpected way to probe magnetic order and we are not aware of any other study having used this before. While we cannot predict uniquely the magnetic order, we can give very precise clues of what can or cannot be compatible with our experiment. At low temperatures (T<90~K), all Co sites are found equivalent, which implies an umbrella structure with a small tilt angle fixed in temperature. Consistent with many other studies of magnetism in \CoSnS \cite{GuguchiaNatCom20,KassemPRB17,ZivkovicPRB22}, we observe an evolution of the magnetic order above 90~K. This is reminiscent of the low frequency line appearing in $\mu$SR in the same temperature range\cite{GuguchiaNatCom20}. However, we exclude a coexisting in-plane AF phase and propose instead a local breaking of the in-plane 3-fold symmetry. A simple way to realize this symmetry breaking is to introduce a tilting of \Bint~in one preferential direction, giving rise to a net in-plane component in a nematic-like fashion, at least on each triangle. This new order will probably be reinforced when the ferromagnetic moment is reduced, as occurs notably for In doping \cite{NeubauerNPJQM22,KassemPhD16,ZivkovicJPSJ23}. This may give new ways to manipulate the magnetic transition and control the formation of the Weyl points \cite{ZhangPRL21}.


\begin{thebibliography}{39}%
	\makeatletter
	\providecommand \@ifxundefined [1]{%
		\@ifx{#1\undefined}
	}%
	\providecommand \@ifnum [1]{%
		\ifnum #1\expandafter \@firstoftwo
		\else \expandafter \@secondoftwo
		\fi
	}%
	\providecommand \@ifx [1]{%
		\ifx #1\expandafter \@firstoftwo
		\else \expandafter \@secondoftwo
		\fi
	}%
	\providecommand \natexlab [1]{#1}%
	\providecommand \enquote  [1]{``#1''}%
	\providecommand \bibnamefont  [1]{#1}%
	\providecommand \bibfnamefont [1]{#1}%
	\providecommand \citenamefont [1]{#1}%
	\providecommand \href@noop [0]{\@secondoftwo}%
	\providecommand \href [0]{\begingroup \@sanitize@url \@href}%
	\providecommand \@href[1]{\@@startlink{#1}\@@href}%
	\providecommand \@@href[1]{\endgroup#1\@@endlink}%
	\providecommand \@sanitize@url [0]{\catcode `\\12\catcode `\$12\catcode
		`\&12\catcode `\#12\catcode `\^12\catcode `\_12\catcode `\%12\relax}%
	\providecommand \@@startlink[1]{}%
	\providecommand \@@endlink[0]{}%
	\providecommand \url  [0]{\begingroup\@sanitize@url \@url }%
	\providecommand \@url [1]{\endgroup\@href {#1}{\urlprefix }}%
	\providecommand \urlprefix  [0]{URL }%
	\providecommand \Eprint [0]{\href }%
	\providecommand \doibase [0]{https://doi.org/}%
	\providecommand \selectlanguage [0]{\@gobble}%
	\providecommand \bibinfo  [0]{\@secondoftwo}%
	\providecommand \bibfield  [0]{\@secondoftwo}%
	\providecommand \translation [1]{[#1]}%
	\providecommand \BibitemOpen [0]{}%
	\providecommand \bibitemStop [0]{}%
	\providecommand \bibitemNoStop [0]{.\EOS\space}%
	\providecommand \EOS [0]{\spacefactor3000\relax}%
	\providecommand \BibitemShut  [1]{\csname bibitem#1\endcsname}%
	\let\auto@bib@innerbib\@empty
	\bibitem [{\citenamefont {Nakatsuji}(2022)}]{Nakatsuji2022}%
	\BibitemOpen
	\bibfield  {author} {\bibinfo {author} {\bibfnamefont {S.}~\bibnamefont
			{Nakatsuji}},\ }\bibfield  {title} {\bibinfo {title} {Topological
			magnets---their basic science and potential applications},\ }\href
	{https://doi.org/10.1007/s43673-022-00046-3} {\bibfield  {journal} {\bibinfo
			{journal} {AAPPS Bulletin}\ }\textbf {\bibinfo {volume} {32}},\ \bibinfo
		{pages} {25} (\bibinfo {year} {2022})}\BibitemShut {NoStop}%
	\bibitem [{\citenamefont {Haldane}(1988)}]{HaldanePRL88}%
	\BibitemOpen
	\bibfield  {author} {\bibinfo {author} {\bibfnamefont {F.~D.~M.}\
			\bibnamefont {Haldane}},\ }\bibfield  {title} {\bibinfo {title} {Model for a
			quantum hall effect without landau levels: Condensed-matter realization of
			the "parity anomaly"},\ }\href {https://doi.org/10.1103/PhysRevLett.61.2015}
	{\bibfield  {journal} {\bibinfo  {journal} {Phys. Rev. Lett.}\ }\textbf
		{\bibinfo {volume} {61}},\ \bibinfo {pages} {2015} (\bibinfo {year}
		{1988})}\BibitemShut {NoStop}%
	\bibitem [{\citenamefont {Guo}\ and\ \citenamefont {Franz}(2009)}]{GuoPRB09}%
	\BibitemOpen
	\bibfield  {author} {\bibinfo {author} {\bibfnamefont {H.-M.}\ \bibnamefont
			{Guo}}\ and\ \bibinfo {author} {\bibfnamefont {M.}~\bibnamefont {Franz}},\
	}\bibfield  {title} {\bibinfo {title} {Topological insulator on the kagome
			lattice},\ }\href {https://doi.org/10.1103/PhysRevB.80.113102} {\bibfield
		{journal} {\bibinfo  {journal} {Phys. Rev. B}\ }\textbf {\bibinfo {volume}
			{80}},\ \bibinfo {pages} {113102} (\bibinfo {year} {2009})}\BibitemShut
	{NoStop}%
	\bibitem [{\citenamefont {Gilmutdinov}\ \emph {et~al.}(2021)\citenamefont
		{Gilmutdinov}, \citenamefont {Sch\"onemann}, \citenamefont {Vignolles},
		\citenamefont {Proust}, \citenamefont {Mukhamedshin}, \citenamefont
		{Balicas},\ and\ \citenamefont {Alloul}}]{GilmutdinovPRB21}%
	\BibitemOpen
	\bibfield  {author} {\bibinfo {author} {\bibfnamefont {I.~F.}\ \bibnamefont
			{Gilmutdinov}}, \bibinfo {author} {\bibfnamefont {R.}~\bibnamefont
			{Sch\"onemann}}, \bibinfo {author} {\bibfnamefont {D.}~\bibnamefont
			{Vignolles}}, \bibinfo {author} {\bibfnamefont {C.}~\bibnamefont {Proust}},
		\bibinfo {author} {\bibfnamefont {I.~R.}\ \bibnamefont {Mukhamedshin}},
		\bibinfo {author} {\bibfnamefont {L.}~\bibnamefont {Balicas}},\ and\ \bibinfo
		{author} {\bibfnamefont {H.}~\bibnamefont {Alloul}},\ }\bibfield  {title}
	{\bibinfo {title} {Interplay between strong correlations and electronic
			topology in the underlying kagome lattice of
			${\mathrm{na}}_{2/3}\mathrm{Co}{\mathrm{o}}_{2}$},\ }\href
	{https://doi.org/10.1103/PhysRevB.104.L201103} {\bibfield  {journal}
		{\bibinfo  {journal} {Phys. Rev. B}\ }\textbf {\bibinfo {volume} {104}},\
		\bibinfo {pages} {L201103} (\bibinfo {year} {2021})}\BibitemShut {NoStop}%
	\bibitem [{\citenamefont {Yin}\ \emph {et~al.}(2022)\citenamefont {Yin},
		\citenamefont {Lian},\ and\ \citenamefont {Hasan}}]{YinHasan22}%
	\BibitemOpen
	\bibfield  {author} {\bibinfo {author} {\bibfnamefont {J.-X.}\ \bibnamefont
			{Yin}}, \bibinfo {author} {\bibfnamefont {B.}~\bibnamefont {Lian}},\ and\
		\bibinfo {author} {\bibfnamefont {M.~Z.}\ \bibnamefont {Hasan}},\ }\bibfield
	{title} {\bibinfo {title} {Topological kagome magnets and superconductors},\
	}\href {https://doi.org/10.1038/s41586-022-05516-0} {\bibfield  {journal}
		{\bibinfo  {journal} {Nature}\ }\textbf {\bibinfo {volume} {612}},\ \bibinfo
		{pages} {647} (\bibinfo {year} {2022})}\BibitemShut {NoStop}%
	\bibitem [{\citenamefont {Nakatsuji}\ \emph {et~al.}(2015)\citenamefont
		{Nakatsuji}, \citenamefont {Kiyohara},\ and\ \citenamefont
		{Higo}}]{NakatsujiNature2015}%
	\BibitemOpen
	\bibfield  {author} {\bibinfo {author} {\bibfnamefont {S.}~\bibnamefont
			{Nakatsuji}}, \bibinfo {author} {\bibfnamefont {N.}~\bibnamefont
			{Kiyohara}},\ and\ \bibinfo {author} {\bibfnamefont {T.}~\bibnamefont
			{Higo}},\ }\bibfield  {title} {\bibinfo {title} {Large anomalous hall effect
			in a non-collinear antiferromagnet at room temperature},\ }\href
	{https://doi.org/10.1038/nature15723} {\bibfield  {journal} {\bibinfo
			{journal} {Nature}\ }\textbf {\bibinfo {volume} {527}},\ \bibinfo {pages}
		{212} (\bibinfo {year} {2015})}\BibitemShut {NoStop}%
	\bibitem [{\citenamefont {Liu}\ \emph {et~al.}(2018)\citenamefont {Liu},
		\citenamefont {Sun}, \citenamefont {Kumar}, \citenamefont {Muechler},
		\citenamefont {Sun}, \citenamefont {Jiao}, \citenamefont {Yang},
		\citenamefont {Liu}, \citenamefont {Liang}, \citenamefont {Xu}, \citenamefont
		{Kroder}, \citenamefont {S{\"u}{\ss}}, \citenamefont {Borrmann},
		\citenamefont {Shekhar}, \citenamefont {Wang}, \citenamefont {Xi},
		\citenamefont {Wang}, \citenamefont {Schnelle}, \citenamefont {Wirth},
		\citenamefont {Chen}, \citenamefont {Goennenwein},\ and\ \citenamefont
		{Felser}}]{LiuNatPhys18}%
	\BibitemOpen
	\bibfield  {author} {\bibinfo {author} {\bibfnamefont {E.}~\bibnamefont
			{Liu}}, \bibinfo {author} {\bibfnamefont {Y.}~\bibnamefont {Sun}}, \bibinfo
		{author} {\bibfnamefont {N.}~\bibnamefont {Kumar}}, \bibinfo {author}
		{\bibfnamefont {L.}~\bibnamefont {Muechler}}, \bibinfo {author}
		{\bibfnamefont {A.}~\bibnamefont {Sun}}, \bibinfo {author} {\bibfnamefont
			{L.}~\bibnamefont {Jiao}}, \bibinfo {author} {\bibfnamefont {S.-Y.}\
			\bibnamefont {Yang}}, \bibinfo {author} {\bibfnamefont {D.}~\bibnamefont
			{Liu}}, \bibinfo {author} {\bibfnamefont {A.}~\bibnamefont {Liang}}, \bibinfo
		{author} {\bibfnamefont {Q.}~\bibnamefont {Xu}}, \bibinfo {author}
		{\bibfnamefont {J.}~\bibnamefont {Kroder}}, \bibinfo {author} {\bibfnamefont
			{V.}~\bibnamefont {S{\"u}{\ss}}}, \bibinfo {author} {\bibfnamefont
			{H.}~\bibnamefont {Borrmann}}, \bibinfo {author} {\bibfnamefont
			{C.}~\bibnamefont {Shekhar}}, \bibinfo {author} {\bibfnamefont
			{Z.}~\bibnamefont {Wang}}, \bibinfo {author} {\bibfnamefont {C.}~\bibnamefont
			{Xi}}, \bibinfo {author} {\bibfnamefont {W.}~\bibnamefont {Wang}}, \bibinfo
		{author} {\bibfnamefont {W.}~\bibnamefont {Schnelle}}, \bibinfo {author}
		{\bibfnamefont {S.}~\bibnamefont {Wirth}}, \bibinfo {author} {\bibfnamefont
			{Y.}~\bibnamefont {Chen}}, \bibinfo {author} {\bibfnamefont {S.~T.~B.}\
			\bibnamefont {Goennenwein}},\ and\ \bibinfo {author} {\bibfnamefont
			{C.}~\bibnamefont {Felser}},\ }\bibfield  {title} {\bibinfo {title} {Giant
			anomalous hall effect in a ferromagnetic kagome-lattice semimetal},\ }\href
	{https://doi.org/10.1038/s41567-018-0234-5} {\bibfield  {journal} {\bibinfo
			{journal} {Nature Physics}\ }\textbf {\bibinfo {volume} {14}},\ \bibinfo
		{pages} {1125} (\bibinfo {year} {2018})}\BibitemShut {NoStop}%
	\bibitem [{\citenamefont {Liu}\ \emph {et~al.}(2019)\citenamefont {Liu},
		\citenamefont {Liang}, \citenamefont {Liu}, \citenamefont {Xu}, \citenamefont
		{Li}, \citenamefont {Chen}, \citenamefont {Pei}, \citenamefont {Shi},
		\citenamefont {Mo}, \citenamefont {Dudin}, \citenamefont {Kim}, \citenamefont
		{Cacho}, \citenamefont {Li}, \citenamefont {Sun}, \citenamefont {Yang},
		\citenamefont {Liu}, \citenamefont {Parkin}, \citenamefont {Felser},\ and\
		\citenamefont {Chen}}]{LiuScience19}%
	\BibitemOpen
	\bibfield  {author} {\bibinfo {author} {\bibfnamefont {D.~F.}\ \bibnamefont
			{Liu}}, \bibinfo {author} {\bibfnamefont {A.~J.}\ \bibnamefont {Liang}},
		\bibinfo {author} {\bibfnamefont {E.~K.}\ \bibnamefont {Liu}}, \bibinfo
		{author} {\bibfnamefont {Q.~N.}\ \bibnamefont {Xu}}, \bibinfo {author}
		{\bibfnamefont {Y.~W.}\ \bibnamefont {Li}}, \bibinfo {author} {\bibfnamefont
			{C.}~\bibnamefont {Chen}}, \bibinfo {author} {\bibfnamefont {D.}~\bibnamefont
			{Pei}}, \bibinfo {author} {\bibfnamefont {W.~J.}\ \bibnamefont {Shi}},
		\bibinfo {author} {\bibfnamefont {S.~K.}\ \bibnamefont {Mo}}, \bibinfo
		{author} {\bibfnamefont {P.}~\bibnamefont {Dudin}}, \bibinfo {author}
		{\bibfnamefont {T.}~\bibnamefont {Kim}}, \bibinfo {author} {\bibfnamefont
			{C.}~\bibnamefont {Cacho}}, \bibinfo {author} {\bibfnamefont
			{G.}~\bibnamefont {Li}}, \bibinfo {author} {\bibfnamefont {Y.}~\bibnamefont
			{Sun}}, \bibinfo {author} {\bibfnamefont {L.~X.}\ \bibnamefont {Yang}},
		\bibinfo {author} {\bibfnamefont {Z.~K.}\ \bibnamefont {Liu}}, \bibinfo
		{author} {\bibfnamefont {S.~S.~P.}\ \bibnamefont {Parkin}}, \bibinfo {author}
		{\bibfnamefont {C.}~\bibnamefont {Felser}},\ and\ \bibinfo {author}
		{\bibfnamefont {Y.~L.}\ \bibnamefont {Chen}},\ }\bibfield  {title} {\bibinfo
		{title} {Magnetic weyl semimetal phase in a kagomé crystal},\ }\href
	{https://doi.org/10.1126/science.aav2873} {\bibfield  {journal} {\bibinfo
			{journal} {Science}\ }\textbf {\bibinfo {volume} {365}},\ \bibinfo {pages}
		{1282} (\bibinfo {year} {2019})},\ \Eprint
	{https://arxiv.org/abs/https://www.science.org/doi/pdf/10.1126/science.aav2873}
	{https://www.science.org/doi/pdf/10.1126/science.aav2873} \BibitemShut
	{NoStop}%
	\bibitem [{\citenamefont {Schnelle}\ \emph {et~al.}(2013)\citenamefont
		{Schnelle}, \citenamefont {Leithe-Jasper}, \citenamefont {Rosner},
		\citenamefont {Schappacher}, \citenamefont {P\"ottgen}, \citenamefont
		{Pielnhofer},\ and\ \citenamefont {Weihrich}}]{SchnellePRB13}%
	\BibitemOpen
	\bibfield  {author} {\bibinfo {author} {\bibfnamefont {W.}~\bibnamefont
			{Schnelle}}, \bibinfo {author} {\bibfnamefont {A.}~\bibnamefont
			{Leithe-Jasper}}, \bibinfo {author} {\bibfnamefont {H.}~\bibnamefont
			{Rosner}}, \bibinfo {author} {\bibfnamefont {F.~M.}\ \bibnamefont
			{Schappacher}}, \bibinfo {author} {\bibfnamefont {R.}~\bibnamefont
			{P\"ottgen}}, \bibinfo {author} {\bibfnamefont {F.}~\bibnamefont
			{Pielnhofer}},\ and\ \bibinfo {author} {\bibfnamefont {R.}~\bibnamefont
			{Weihrich}},\ }\bibfield  {title} {\bibinfo {title} {Ferromagnetic ordering
			and half-metallic state of sn${}_{2}$co${}_{3}$s${}_{2}$ with the
			shandite-type structure},\ }\href
	{https://doi.org/10.1103/PhysRevB.88.144404} {\bibfield  {journal} {\bibinfo
			{journal} {Phys. Rev. B}\ }\textbf {\bibinfo {volume} {88}},\ \bibinfo
		{pages} {144404} (\bibinfo {year} {2013})}\BibitemShut {NoStop}%
	\bibitem [{\citenamefont {Lohani}\ \emph {et~al.}(2023)\citenamefont {Lohani},
		\citenamefont {Foulquier}, \citenamefont {Le~F\`evre}, \citenamefont
		{Bertran}, \citenamefont {Colson}, \citenamefont {Forget},\ and\
		\citenamefont {Brouet}}]{LohaniPRB23}%
	\BibitemOpen
	\bibfield  {author} {\bibinfo {author} {\bibfnamefont {H.}~\bibnamefont
			{Lohani}}, \bibinfo {author} {\bibfnamefont {P.}~\bibnamefont {Foulquier}},
		\bibinfo {author} {\bibfnamefont {P.}~\bibnamefont {Le~F\`evre}}, \bibinfo
		{author} {\bibfnamefont {F.~m.~c.}\ \bibnamefont {Bertran}}, \bibinfo
		{author} {\bibfnamefont {D.}~\bibnamefont {Colson}}, \bibinfo {author}
		{\bibfnamefont {A.}~\bibnamefont {Forget}},\ and\ \bibinfo {author}
		{\bibfnamefont {V.}~\bibnamefont {Brouet}},\ }\bibfield  {title} {\bibinfo
		{title} {Electronic structure evolution of the magnetic weyl semimetal
			${\mathrm{co}}_{3}{\mathrm{sn}}_{2}{\mathrm{s}}_{2}$ with hole and electron
			doping},\ }\href {https://doi.org/10.1103/PhysRevB.107.245119} {\bibfield
		{journal} {\bibinfo  {journal} {Phys. Rev. B}\ }\textbf {\bibinfo {volume}
			{107}},\ \bibinfo {pages} {245119} (\bibinfo {year} {2023})}\BibitemShut
	{NoStop}%
	\bibitem [{\citenamefont {Mendels}\ and\ \citenamefont
		{Bert}(2016)}]{Mendels16}%
	\BibitemOpen
	\bibfield  {author} {\bibinfo {author} {\bibfnamefont {P.}~\bibnamefont
			{Mendels}}\ and\ \bibinfo {author} {\bibfnamefont {F.}~\bibnamefont {Bert}},\
	}\bibfield  {title} {\bibinfo {title} {Quantum kagome frustrated
			antiferromagnets: One route to quantum spin liquids},\ }\href
	{https://doi.org/https://doi.org/10.1016/j.crhy.2015.12.001} {\bibfield
		{journal} {\bibinfo  {journal} {Comptes Rendus Physique}\ }\textbf {\bibinfo
			{volume} {17}},\ \bibinfo {pages} {455} (\bibinfo {year} {2016})},\ \bibinfo
	{note} {physique de la mati\`{e}re condens\'{e}e au XXIe si\`{e}cle:
		l'h\'{e}ritage de Jacques Friedel}\BibitemShut {NoStop}%
	\bibitem [{\citenamefont {Solovyev}\ \emph {et~al.}(2022)\citenamefont
		{Solovyev}, \citenamefont {Nikolaev}, \citenamefont {Ushakov}, \citenamefont
		{Irkhin}, \citenamefont {Tanaka},\ and\ \citenamefont
		{Streltsov}}]{SolovyevPRB22}%
	\BibitemOpen
	\bibfield  {author} {\bibinfo {author} {\bibfnamefont {I.~V.}\ \bibnamefont
			{Solovyev}}, \bibinfo {author} {\bibfnamefont {S.~A.}\ \bibnamefont
			{Nikolaev}}, \bibinfo {author} {\bibfnamefont {A.~V.}\ \bibnamefont
			{Ushakov}}, \bibinfo {author} {\bibfnamefont {V.~Y.}\ \bibnamefont {Irkhin}},
		\bibinfo {author} {\bibfnamefont {A.}~\bibnamefont {Tanaka}},\ and\ \bibinfo
		{author} {\bibfnamefont {S.~V.}\ \bibnamefont {Streltsov}},\ }\bibfield
	{title} {\bibinfo {title} {Microscopic origins and stability of
			ferromagnetism in ${\mathrm{co}}_{3}{\mathrm{sn}}_{2}{\mathrm{s}}_{2}$},\
	}\href {https://doi.org/10.1103/PhysRevB.105.014415} {\bibfield  {journal}
		{\bibinfo  {journal} {Phys. Rev. B}\ }\textbf {\bibinfo {volume} {105}},\
		\bibinfo {pages} {014415} (\bibinfo {year} {2022})}\BibitemShut {NoStop}%
	\bibitem [{\citenamefont {Guguchia}\ \emph {et~al.}(2020)\citenamefont
		{Guguchia}, \citenamefont {Verezhak}, \citenamefont {Gawryluk}, \citenamefont
		{Tsirkin}, \citenamefont {Yin}, \citenamefont {Belopolski}, \citenamefont
		{Zhou}, \citenamefont {Simutis}, \citenamefont {Zhang}, \citenamefont
		{Cochran}, \citenamefont {Chang}, \citenamefont {Pomjakushina}, \citenamefont
		{Keller}, \citenamefont {Skrzeczkowska}, \citenamefont {Wang}, \citenamefont
		{Lei}, \citenamefont {Khasanov}, \citenamefont {Amato}, \citenamefont {Jia},
		\citenamefont {Neupert}, \citenamefont {Luetkens},\ and\ \citenamefont
		{Hasan}}]{GuguchiaNatCom20}%
	\BibitemOpen
	\bibfield  {author} {\bibinfo {author} {\bibfnamefont {Z.}~\bibnamefont
			{Guguchia}}, \bibinfo {author} {\bibfnamefont {J.~A.~T.}\ \bibnamefont
			{Verezhak}}, \bibinfo {author} {\bibfnamefont {D.~J.}\ \bibnamefont
			{Gawryluk}}, \bibinfo {author} {\bibfnamefont {S.~S.}\ \bibnamefont
			{Tsirkin}}, \bibinfo {author} {\bibfnamefont {J.-X.}\ \bibnamefont {Yin}},
		\bibinfo {author} {\bibfnamefont {I.}~\bibnamefont {Belopolski}}, \bibinfo
		{author} {\bibfnamefont {H.}~\bibnamefont {Zhou}}, \bibinfo {author}
		{\bibfnamefont {G.}~\bibnamefont {Simutis}}, \bibinfo {author} {\bibfnamefont
			{S.-S.}\ \bibnamefont {Zhang}}, \bibinfo {author} {\bibfnamefont {T.~A.}\
			\bibnamefont {Cochran}}, \bibinfo {author} {\bibfnamefont {G.}~\bibnamefont
			{Chang}}, \bibinfo {author} {\bibfnamefont {E.}~\bibnamefont {Pomjakushina}},
		\bibinfo {author} {\bibfnamefont {L.}~\bibnamefont {Keller}}, \bibinfo
		{author} {\bibfnamefont {Z.}~\bibnamefont {Skrzeczkowska}}, \bibinfo {author}
		{\bibfnamefont {Q.}~\bibnamefont {Wang}}, \bibinfo {author} {\bibfnamefont
			{H.~C.}\ \bibnamefont {Lei}}, \bibinfo {author} {\bibfnamefont
			{R.}~\bibnamefont {Khasanov}}, \bibinfo {author} {\bibfnamefont
			{A.}~\bibnamefont {Amato}}, \bibinfo {author} {\bibfnamefont
			{S.}~\bibnamefont {Jia}}, \bibinfo {author} {\bibfnamefont {T.}~\bibnamefont
			{Neupert}}, \bibinfo {author} {\bibfnamefont {H.}~\bibnamefont {Luetkens}},\
		and\ \bibinfo {author} {\bibfnamefont {M.~Z.}\ \bibnamefont {Hasan}},\
	}\bibfield  {title} {\bibinfo {title} {Tunable anomalous hall conductivity
			through volume-wise magnetic competition in a topological kagome magnet},\
	}\href {https://doi.org/10.1038/s41467-020-14325-w} {\bibfield  {journal}
		{\bibinfo  {journal} {Nature Communications}\ }\textbf {\bibinfo {volume}
			{11}},\ \bibinfo {pages} {559} (\bibinfo {year} {2020})}\BibitemShut
	{NoStop}%
	\bibitem [{\citenamefont {Soh}\ \emph {et~al.}(2022)\citenamefont {Soh},
		\citenamefont {Yi}, \citenamefont {Zivkovic}, \citenamefont {Qureshi},
		\citenamefont {Stunault}, \citenamefont {Ouladdiaf}, \citenamefont
		{Rodr\'{\i}guez-Velamaz\'an}, \citenamefont {Shi}, \citenamefont {Ronnow},\
		and\ \citenamefont {Boothroyd}}]{SohBoothroydPRB22}%
	\BibitemOpen
	\bibfield  {author} {\bibinfo {author} {\bibfnamefont {J.-R.}\ \bibnamefont
			{Soh}}, \bibinfo {author} {\bibfnamefont {C.}~\bibnamefont {Yi}}, \bibinfo
		{author} {\bibfnamefont {I.}~\bibnamefont {Zivkovic}}, \bibinfo {author}
		{\bibfnamefont {N.}~\bibnamefont {Qureshi}}, \bibinfo {author} {\bibfnamefont
			{A.}~\bibnamefont {Stunault}}, \bibinfo {author} {\bibfnamefont
			{B.}~\bibnamefont {Ouladdiaf}}, \bibinfo {author} {\bibfnamefont {J.~A.}\
			\bibnamefont {Rodr\'{\i}guez-Velamaz\'an}}, \bibinfo {author} {\bibfnamefont
			{Y.}~\bibnamefont {Shi}}, \bibinfo {author} {\bibfnamefont {H.~M.}\
			\bibnamefont {Ronnow}},\ and\ \bibinfo {author} {\bibfnamefont {A.~T.}\
			\bibnamefont {Boothroyd}},\ }\bibfield  {title} {\bibinfo {title} {Magnetic
			structure of the topological semimetal
			${\mathrm{co}}_{3}{\mathrm{sn}}_{2}{\mathrm{s}}_{2}$},\ }\href
	{https://doi.org/10.1103/PhysRevB.105.094435} {\bibfield  {journal} {\bibinfo
			{journal} {Phys. Rev. B}\ }\textbf {\bibinfo {volume} {105}},\ \bibinfo
		{pages} {094435} (\bibinfo {year} {2022})}\BibitemShut {NoStop}%
	\bibitem [{\citenamefont {Neubauer}\ \emph {et~al.}(2022)\citenamefont
		{Neubauer}, \citenamefont {Ye}, \citenamefont {Shi}, \citenamefont
		{Malinowski}, \citenamefont {Gao}, \citenamefont {Taddei}, \citenamefont
		{Bourges}, \citenamefont {Ivanov}, \citenamefont {Chu},\ and\ \citenamefont
		{Dai}}]{NeubauerNPJQM22}%
	\BibitemOpen
	\bibfield  {author} {\bibinfo {author} {\bibfnamefont {K.~J.}\ \bibnamefont
			{Neubauer}}, \bibinfo {author} {\bibfnamefont {F.}~\bibnamefont {Ye}},
		\bibinfo {author} {\bibfnamefont {Y.}~\bibnamefont {Shi}}, \bibinfo {author}
		{\bibfnamefont {P.}~\bibnamefont {Malinowski}}, \bibinfo {author}
		{\bibfnamefont {B.}~\bibnamefont {Gao}}, \bibinfo {author} {\bibfnamefont
			{K.~M.}\ \bibnamefont {Taddei}}, \bibinfo {author} {\bibfnamefont
			{P.}~\bibnamefont {Bourges}}, \bibinfo {author} {\bibfnamefont
			{A.}~\bibnamefont {Ivanov}}, \bibinfo {author} {\bibfnamefont {J.-H.}\
			\bibnamefont {Chu}},\ and\ \bibinfo {author} {\bibfnamefont {P.}~\bibnamefont
			{Dai}},\ }\bibfield  {title} {\bibinfo {title} {Spin structure and dynamics
			of the topological semimetal co3sn2-xinxs2},\ }\href
	{https://doi.org/10.1038/s41535-022-00523-w} {\bibfield  {journal} {\bibinfo
			{journal} {npj Quantum Materials}\ }\textbf {\bibinfo {volume} {7}},\
		\bibinfo {pages} {112} (\bibinfo {year} {2022})}\BibitemShut {NoStop}%
	\bibitem [{\citenamefont {Kassem}(2016)}]{KassemPhD16}%
	\BibitemOpen
	\bibfield  {author} {\bibinfo {author} {\bibfnamefont {M.~A.}\ \bibnamefont
			{Kassem}},\ }\href@noop {} {\bibfield  {journal} {\bibinfo  {journal} {Ph.D.
				Dissertation, Kyoto University}\ } (\bibinfo {year} {2016})}\BibitemShut
	{NoStop}%
	\bibitem [{\citenamefont {Kassem}\ \emph {et~al.}(2017)\citenamefont {Kassem},
		\citenamefont {Tabata}, \citenamefont {Waki},\ and\ \citenamefont
		{Nakamura}}]{KassemPRB17}%
	\BibitemOpen
	\bibfield  {author} {\bibinfo {author} {\bibfnamefont {M.~A.}\ \bibnamefont
			{Kassem}}, \bibinfo {author} {\bibfnamefont {Y.}~\bibnamefont {Tabata}},
		\bibinfo {author} {\bibfnamefont {T.}~\bibnamefont {Waki}},\ and\ \bibinfo
		{author} {\bibfnamefont {H.}~\bibnamefont {Nakamura}},\ }\bibfield  {title}
	{\bibinfo {title} {Low-field anomalous magnetic phase in the kagome-lattice
			shandite
			$\mathrm{C}{\mathrm{o}}_{3}\mathrm{S}{\mathrm{n}}_{2}{\mathrm{s}}_{2}$},\
	}\href {https://doi.org/10.1103/PhysRevB.96.014429} {\bibfield  {journal}
		{\bibinfo  {journal} {Phys. Rev. B}\ }\textbf {\bibinfo {volume} {96}},\
		\bibinfo {pages} {014429} (\bibinfo {year} {2017})}\BibitemShut {NoStop}%
	\bibitem [{\citenamefont {Sugawara}\ \emph {et~al.}(2019)\citenamefont
		{Sugawara}, \citenamefont {Akashi}, \citenamefont {Kassem}, \citenamefont
		{Tabata}, \citenamefont {Waki},\ and\ \citenamefont
		{Nakamura}}]{SugawaraPRM19}%
	\BibitemOpen
	\bibfield  {author} {\bibinfo {author} {\bibfnamefont {A.}~\bibnamefont
			{Sugawara}}, \bibinfo {author} {\bibfnamefont {T.}~\bibnamefont {Akashi}},
		\bibinfo {author} {\bibfnamefont {M.~A.}\ \bibnamefont {Kassem}}, \bibinfo
		{author} {\bibfnamefont {Y.}~\bibnamefont {Tabata}}, \bibinfo {author}
		{\bibfnamefont {T.}~\bibnamefont {Waki}},\ and\ \bibinfo {author}
		{\bibfnamefont {H.}~\bibnamefont {Nakamura}},\ }\bibfield  {title} {\bibinfo
		{title} {Magnetic domain structure within half-metallic ferromagnetic kagome
			compound
			$\mathrm{C}{\mathrm{o}}_{3}\mathrm{S}{\mathrm{n}}_{2}{\mathrm{s}}_{2}$},\
	}\href {https://doi.org/10.1103/PhysRevMaterials.3.104421} {\bibfield
		{journal} {\bibinfo  {journal} {Phys. Rev. Mater.}\ }\textbf {\bibinfo
			{volume} {3}},\ \bibinfo {pages} {104421} (\bibinfo {year}
		{2019})}\BibitemShut {NoStop}%
	\bibitem [{\citenamefont {Lee}\ \emph {et~al.}(2022)\citenamefont {Lee},
		\citenamefont {Vir}, \citenamefont {Manna}, \citenamefont {Shekhar},
		\citenamefont {Moore}, \citenamefont {Kastner}, \citenamefont {Felser},\ and\
		\citenamefont {Orenstein}}]{LeeNatCom22}%
	\BibitemOpen
	\bibfield  {author} {\bibinfo {author} {\bibfnamefont {C.}~\bibnamefont
			{Lee}}, \bibinfo {author} {\bibfnamefont {P.}~\bibnamefont {Vir}}, \bibinfo
		{author} {\bibfnamefont {K.}~\bibnamefont {Manna}}, \bibinfo {author}
		{\bibfnamefont {C.}~\bibnamefont {Shekhar}}, \bibinfo {author} {\bibfnamefont
			{J.~E.}\ \bibnamefont {Moore}}, \bibinfo {author} {\bibfnamefont {M.~A.}\
			\bibnamefont {Kastner}}, \bibinfo {author} {\bibfnamefont {C.}~\bibnamefont
			{Felser}},\ and\ \bibinfo {author} {\bibfnamefont {J.}~\bibnamefont
			{Orenstein}},\ }\bibfield  {title} {\bibinfo {title} {Observation of a phase
			transition within the domain walls of ferromagnetic co3sn2s2},\ }\href
	{https://doi.org/10.1038/s41467-022-30460-y} {\bibfield  {journal} {\bibinfo
			{journal} {Nature Communications}\ }\textbf {\bibinfo {volume} {13}},\
		\bibinfo {pages} {3000} (\bibinfo {year} {2022})}\BibitemShut {NoStop}%
	\bibitem [{\citenamefont {Zivkovic}\ \emph {et~al.}(2022)\citenamefont
		{Zivkovic}, \citenamefont {Yadav}, \citenamefont {Soh}, \citenamefont {Yi},
		\citenamefont {Shi}, \citenamefont {Yazyev},\ and\ \citenamefont
		{Ronnow}}]{ZivkovicPRB22}%
	\BibitemOpen
	\bibfield  {author} {\bibinfo {author} {\bibfnamefont {I.}~\bibnamefont
			{Zivkovic}}, \bibinfo {author} {\bibfnamefont {R.}~\bibnamefont {Yadav}},
		\bibinfo {author} {\bibfnamefont {J.-R.}\ \bibnamefont {Soh}}, \bibinfo
		{author} {\bibfnamefont {C.}~\bibnamefont {Yi}}, \bibinfo {author}
		{\bibfnamefont {Y.}~\bibnamefont {Shi}}, \bibinfo {author} {\bibfnamefont
			{O.~V.}\ \bibnamefont {Yazyev}},\ and\ \bibinfo {author} {\bibfnamefont
			{H.~M.}\ \bibnamefont {Ronnow}},\ }\bibfield  {title} {\bibinfo {title}
		{Unraveling the origin of the peculiar transition in the magnetically ordered
			phase of the weyl semimetal
			${\mathrm{co}}_{3}{\mathrm{sn}}_{2}{\mathrm{s}}_{2}$},\ }\href
	{https://doi.org/10.1103/PhysRevB.106.L180403} {\bibfield  {journal}
		{\bibinfo  {journal} {Phys. Rev. B}\ }\textbf {\bibinfo {volume} {106}},\
		\bibinfo {pages} {L180403} (\bibinfo {year} {2022})}\BibitemShut {NoStop}%
	\bibitem [{\citenamefont {Abragam}(1961)}]{Abragam}%
	\BibitemOpen
	\bibfield  {author} {\bibinfo {author} {\bibfnamefont {A.}~\bibnamefont
			{Abragam}},\ }\href@noop {} {\emph {\bibinfo {title} {The Principles of
				nuclear magnetism}}}\ (\bibinfo  {publisher} {Oxford: Clarendon Press,
		London},\ \bibinfo {year} {1961})\BibitemShut {NoStop}%
	\bibitem [{sup()}]{sup}%
	\BibitemOpen
	\href@noop {} {\bibinfo  {journal} {See supplementary information (including
			references \cite{Clark,Bussandri})}\ }\BibitemShut {NoStop}%
	\bibitem [{\citenamefont {Turov}\ and\ \citenamefont
		{Petrov}(1972)}]{TurovPetrov}%
	\BibitemOpen
	\bibfield  {journal} {  }\bibfield  {author} {\bibinfo {author} {\bibfnamefont
			{E.~A.}\ \bibnamefont {Turov}}\ and\ \bibinfo {author} {\bibfnamefont
			{M.~P.}\ \bibnamefont {Petrov}},\ }\href@noop {} {\emph {\bibinfo {title}
			{NMR in Ferro- and Antiferromagnets}}}\ (\bibinfo  {publisher} {Halsted
		Press},\ \bibinfo {year} {1972})\BibitemShut {NoStop}%
	\bibitem [{\citenamefont {Mukhamedshin}\ and\ \citenamefont
		{Alloul}(2011)}]{H67_CoNMR}%
	\BibitemOpen
	\bibfield  {author} {\bibinfo {author} {\bibfnamefont {I.~R.}\ \bibnamefont
			{Mukhamedshin}}\ and\ \bibinfo {author} {\bibfnamefont {H.}~\bibnamefont
			{Alloul}},\ }\bibfield  {title} {\bibinfo {title} {$^{59}$co nmr evidence for
			charge and orbital order in the kagome-like structure of
			na$_{2/3}$coo$_{2}$},\ }\href {https://doi.org/10.1103/PhysRevB.84.155112}
	{\bibfield  {journal} {\bibinfo  {journal} {Phys. Rev. B}\ }\textbf {\bibinfo
			{volume} {84}},\ \bibinfo {pages} {155112} (\bibinfo {year}
		{2011})}\BibitemShut {NoStop}%
	\bibitem [{\citenamefont {Mukhamedshin}\ and\ \citenamefont
		{Alloul}(2015)}]{PhysicaB2015}%
	\BibitemOpen
	\bibfield  {author} {\bibinfo {author} {\bibfnamefont {I.~R.}\ \bibnamefont
			{Mukhamedshin}}\ and\ \bibinfo {author} {\bibfnamefont {H.}~\bibnamefont
			{Alloul}},\ }\bibfield  {title} {\bibinfo {title} {{N}a order and {C}o charge
			disproportionation in {N}a$_{x}${C}o{O}$_{2}$},\ }\href@noop {} {\bibfield
		{journal} {\bibinfo  {journal} {Physica B: Condensed Matter}\ }\textbf
		{\bibinfo {volume} {460}},\ \bibinfo {pages} {58 } (\bibinfo {year}
		{2015})}\BibitemShut {NoStop}%
	\bibitem [{\citenamefont {Huang}\ \emph {et~al.}(2022)\citenamefont {Huang},
		\citenamefont {Zheng}, \citenamefont {Lin}, \citenamefont {Guo},
		\citenamefont {Wang}, \citenamefont {Zhang}, \citenamefont {Zhang},
		\citenamefont {Sun}, \citenamefont {Wang}, \citenamefont {Weng},
		\citenamefont {Li}, \citenamefont {Wu}, \citenamefont {Chen},\ and\
		\citenamefont {Zeng}}]{HuangPRL22}%
	\BibitemOpen
	\bibfield  {author} {\bibinfo {author} {\bibfnamefont {H.}~\bibnamefont
			{Huang}}, \bibinfo {author} {\bibfnamefont {L.}~\bibnamefont {Zheng}},
		\bibinfo {author} {\bibfnamefont {Z.}~\bibnamefont {Lin}}, \bibinfo {author}
		{\bibfnamefont {X.}~\bibnamefont {Guo}}, \bibinfo {author} {\bibfnamefont
			{S.}~\bibnamefont {Wang}}, \bibinfo {author} {\bibfnamefont {S.}~\bibnamefont
			{Zhang}}, \bibinfo {author} {\bibfnamefont {C.}~\bibnamefont {Zhang}},
		\bibinfo {author} {\bibfnamefont {Z.}~\bibnamefont {Sun}}, \bibinfo {author}
		{\bibfnamefont {Z.}~\bibnamefont {Wang}}, \bibinfo {author} {\bibfnamefont
			{H.}~\bibnamefont {Weng}}, \bibinfo {author} {\bibfnamefont {L.}~\bibnamefont
			{Li}}, \bibinfo {author} {\bibfnamefont {T.}~\bibnamefont {Wu}}, \bibinfo
		{author} {\bibfnamefont {X.}~\bibnamefont {Chen}},\ and\ \bibinfo {author}
		{\bibfnamefont {C.}~\bibnamefont {Zeng}},\ }\bibfield  {title} {\bibinfo
		{title} {Flat-band-induced anomalous anisotropic charge transport and orbital
			magnetism in kagome metal cosn},\ }\href
	{https://doi.org/10.1103/PhysRevLett.128.096601} {\bibfield  {journal}
		{\bibinfo  {journal} {Phys. Rev. Lett.}\ }\textbf {\bibinfo {volume} {128}},\
		\bibinfo {pages} {096601} (\bibinfo {year} {2022})}\BibitemShut {NoStop}%
	\bibitem [{\citenamefont {Storn}\ and\ \citenamefont
		{Price}(1997)}]{storn_price}%
	\BibitemOpen
	\bibfield  {author} {\bibinfo {author} {\bibfnamefont {R.}~\bibnamefont
			{Storn}}\ and\ \bibinfo {author} {\bibfnamefont {K.}~\bibnamefont {Price}},\
	}\bibfield  {title} {\bibinfo {title} {Differential evolution--a simple and
			efficient heuristic for global optimization over continuous spaces},\
	}\href@noop {} {\bibfield  {journal} {\bibinfo  {journal} {Journal of global
				optimization}\ }\textbf {\bibinfo {volume} {11}},\ \bibinfo {pages} {341}
		(\bibinfo {year} {1997})}\BibitemShut {NoStop}%
	\bibitem [{\citenamefont {Wormington}\ \emph {et~al.}(1999)\citenamefont
		{Wormington}, \citenamefont {Panaccione}, \citenamefont {Matney},\ and\
		\citenamefont {Bowen}}]{wormington}%
	\BibitemOpen
	\bibfield  {author} {\bibinfo {author} {\bibfnamefont {M.}~\bibnamefont
			{Wormington}}, \bibinfo {author} {\bibfnamefont {C.}~\bibnamefont
			{Panaccione}}, \bibinfo {author} {\bibfnamefont {K.~M.}\ \bibnamefont
			{Matney}},\ and\ \bibinfo {author} {\bibfnamefont {D.~K.}\ \bibnamefont
			{Bowen}},\ }\bibfield  {title} {\bibinfo {title} {Characterization of
			structures from x-ray scattering data using genetic algorithms},\ }\href@noop
	{} {\bibfield  {journal} {\bibinfo  {journal} {Philosophical Transactions of
				the Royal Society of London, Series A}\ }\textbf {\bibinfo {volume} {357}},\
		\bibinfo {pages} {2827} (\bibinfo {year} {1999})}\BibitemShut {NoStop}%
	\bibitem [{\citenamefont {Das}\ and\ \citenamefont
		{Suganthan}(2010)}]{das2010de}%
	\BibitemOpen
	\bibfield  {author} {\bibinfo {author} {\bibfnamefont {S.}~\bibnamefont
			{Das}}\ and\ \bibinfo {author} {\bibfnamefont {P.~N.}\ \bibnamefont
			{Suganthan}},\ }\bibfield  {title} {\bibinfo {title} {Differential evolution:
			A survey of the state-of-the-art},\ }\href@noop {} {\bibfield  {journal}
		{\bibinfo  {journal} {IEEE transactions on evolutionary computation}\
		}\textbf {\bibinfo {volume} {15}},\ \bibinfo {pages} {4} (\bibinfo {year}
		{2010})}\BibitemShut {NoStop}%
	\bibitem [{\citenamefont {Xu}\ \emph {et~al.}(2018)\citenamefont {Xu},
		\citenamefont {Liu}, \citenamefont {Shi}, \citenamefont {Muechler},
		\citenamefont {Gayles}, \citenamefont {Felser},\ and\ \citenamefont
		{Sun}}]{XuFelserPRB18}%
	\BibitemOpen
	\bibfield  {author} {\bibinfo {author} {\bibfnamefont {Q.}~\bibnamefont
			{Xu}}, \bibinfo {author} {\bibfnamefont {E.}~\bibnamefont {Liu}}, \bibinfo
		{author} {\bibfnamefont {W.}~\bibnamefont {Shi}}, \bibinfo {author}
		{\bibfnamefont {L.}~\bibnamefont {Muechler}}, \bibinfo {author}
		{\bibfnamefont {J.}~\bibnamefont {Gayles}}, \bibinfo {author} {\bibfnamefont
			{C.}~\bibnamefont {Felser}},\ and\ \bibinfo {author} {\bibfnamefont
			{Y.}~\bibnamefont {Sun}},\ }\bibfield  {title} {\bibinfo {title} {Topological
			surface fermi arcs in the magnetic weyl semimetal
			${\mathrm{co}}_{3}{\mathrm{sn}}_{2}{\mathrm{s}}_{2}$},\ }\href
	{https://doi.org/10.1103/PhysRevB.97.235416} {\bibfield  {journal} {\bibinfo
			{journal} {Phys. Rev. B}\ }\textbf {\bibinfo {volume} {97}},\ \bibinfo
		{pages} {235416} (\bibinfo {year} {2018})}\BibitemShut {NoStop}%
	\bibitem [{\citenamefont {Man}(1995)}]{ManPRB}%
	\BibitemOpen
	\bibfield  {author} {\bibinfo {author} {\bibfnamefont {P.~P.}\ \bibnamefont
			{Man}},\ }\bibfield  {title} {\bibinfo {title} {Numerical analysis of hahn
			echoes in solids},\ }\href@noop {} {\bibfield  {journal} {\bibinfo  {journal}
			{Phys. Rev. B}\ }\textbf {\bibinfo {volume} {52}},\ \bibinfo {pages} {9418}
		(\bibinfo {year} {1995})}\BibitemShut {NoStop}%
	\bibitem [{\citenamefont {Zhang}\ \emph {et~al.}(2022)\citenamefont {Zhang},
		\citenamefont {Zhang}, \citenamefont {Matsuda}, \citenamefont {Garlea},
		\citenamefont {Yan}, \citenamefont {McGuire}, \citenamefont {Tennant},\ and\
		\citenamefont {Okamoto}}]{ZhangJACS22}%
	\BibitemOpen
	\bibfield  {author} {\bibinfo {author} {\bibfnamefont {Q.}~\bibnamefont
			{Zhang}}, \bibinfo {author} {\bibfnamefont {Y.}~\bibnamefont {Zhang}},
		\bibinfo {author} {\bibfnamefont {M.}~\bibnamefont {Matsuda}}, \bibinfo
		{author} {\bibfnamefont {V.~O.}\ \bibnamefont {Garlea}}, \bibinfo {author}
		{\bibfnamefont {J.}~\bibnamefont {Yan}}, \bibinfo {author} {\bibfnamefont
			{M.~A.}\ \bibnamefont {McGuire}}, \bibinfo {author} {\bibfnamefont {D.~A.}\
			\bibnamefont {Tennant}},\ and\ \bibinfo {author} {\bibfnamefont
			{S.}~\bibnamefont {Okamoto}},\ }\bibfield  {title} {\bibinfo {title} {Hidden
			local symmetry breaking in a kagome-lattice magnetic weyl semimetal},\ }\href
	{https://doi.org/10.1021/jacs.2c05665} {\bibfield  {journal} {\bibinfo
			{journal} {Journal of the American Chemical Society}\ }\textbf {\bibinfo
			{volume} {144}},\ \bibinfo {pages} {14339} (\bibinfo {year}
		{2022})}\BibitemShut {NoStop}%
	\bibitem [{\citenamefont {Gainov}\ \emph {et~al.}(2009)\citenamefont {Gainov},
		\citenamefont {Dooglav}, \citenamefont {Pen'kov}, \citenamefont
		{Mukhamedshin}, \citenamefont {Mozgova}, \citenamefont {Evlampiev},\ and\
		\citenamefont {Bryzgalov}}]{CuS_PRB09}%
	\BibitemOpen
	\bibfield  {author} {\bibinfo {author} {\bibfnamefont {R.~R.}\ \bibnamefont
			{Gainov}}, \bibinfo {author} {\bibfnamefont {A.~V.}\ \bibnamefont {Dooglav}},
		\bibinfo {author} {\bibfnamefont {I.~N.}\ \bibnamefont {Pen'kov}}, \bibinfo
		{author} {\bibfnamefont {I.~R.}\ \bibnamefont {Mukhamedshin}}, \bibinfo
		{author} {\bibfnamefont {N.~N.}\ \bibnamefont {Mozgova}}, \bibinfo {author}
		{\bibfnamefont {I.~A.}\ \bibnamefont {Evlampiev}},\ and\ \bibinfo {author}
		{\bibfnamefont {I.~A.}\ \bibnamefont {Bryzgalov}},\ }\bibfield  {title}
	{\bibinfo {title} {Phase transition and anomalous electronic behavior in the
			layered superconductor cus probed by nqr},\ }\href
	{https://doi.org/10.1103/PhysRevB.79.075115} {\bibfield  {journal} {\bibinfo
			{journal} {Phys. Rev. B}\ }\textbf {\bibinfo {volume} {79}},\ \bibinfo
		{pages} {075115} (\bibinfo {year} {2009})}\BibitemShut {NoStop}%
	\bibitem [{\citenamefont {Zhang}\ \emph {et~al.}(2021)\citenamefont {Zhang},
		\citenamefont {Okamoto}, \citenamefont {Samolyuk}, \citenamefont {Stone},
		\citenamefont {Kolesnikov}, \citenamefont {Xue}, \citenamefont {Yan},
		\citenamefont {McGuire}, \citenamefont {Mandrus},\ and\ \citenamefont
		{Tennant}}]{ZhangPRL21}%
	\BibitemOpen
	\bibfield  {author} {\bibinfo {author} {\bibfnamefont {Q.}~\bibnamefont
			{Zhang}}, \bibinfo {author} {\bibfnamefont {S.}~\bibnamefont {Okamoto}},
		\bibinfo {author} {\bibfnamefont {G.~D.}\ \bibnamefont {Samolyuk}}, \bibinfo
		{author} {\bibfnamefont {M.~B.}\ \bibnamefont {Stone}}, \bibinfo {author}
		{\bibfnamefont {A.~I.}\ \bibnamefont {Kolesnikov}}, \bibinfo {author}
		{\bibfnamefont {R.}~\bibnamefont {Xue}}, \bibinfo {author} {\bibfnamefont
			{J.}~\bibnamefont {Yan}}, \bibinfo {author} {\bibfnamefont {M.~A.}\
			\bibnamefont {McGuire}}, \bibinfo {author} {\bibfnamefont {D.}~\bibnamefont
			{Mandrus}},\ and\ \bibinfo {author} {\bibfnamefont {D.~A.}\ \bibnamefont
			{Tennant}},\ }\bibfield  {title} {\bibinfo {title} {Unusual exchange
			couplings and intermediate temperature weyl state in
			${\mathrm{co}}_{3}{\mathrm{sn}}_{2}{\mathrm{s}}_{2}$},\ }\href
	{https://doi.org/10.1103/PhysRevLett.127.117201} {\bibfield  {journal}
		{\bibinfo  {journal} {Phys. Rev. Lett.}\ }\textbf {\bibinfo {volume} {127}},\
		\bibinfo {pages} {117201} (\bibinfo {year} {2021})}\BibitemShut {NoStop}%
	\bibitem [{\citenamefont {Lachman}\ \emph {et~al.}(2020)\citenamefont
		{Lachman}, \citenamefont {Murphy}, \citenamefont {Maksimovic}, \citenamefont
		{Kealhofer}, \citenamefont {Haley}, \citenamefont {McDonald}, \citenamefont
		{Long},\ and\ \citenamefont {Analytis}}]{LachmanAnalytisNatCom20}%
	\BibitemOpen
	\bibfield  {author} {\bibinfo {author} {\bibfnamefont {E.}~\bibnamefont
			{Lachman}}, \bibinfo {author} {\bibfnamefont {R.~A.}\ \bibnamefont {Murphy}},
		\bibinfo {author} {\bibfnamefont {N.}~\bibnamefont {Maksimovic}}, \bibinfo
		{author} {\bibfnamefont {R.}~\bibnamefont {Kealhofer}}, \bibinfo {author}
		{\bibfnamefont {S.}~\bibnamefont {Haley}}, \bibinfo {author} {\bibfnamefont
			{R.~D.}\ \bibnamefont {McDonald}}, \bibinfo {author} {\bibfnamefont {J.~R.}\
			\bibnamefont {Long}},\ and\ \bibinfo {author} {\bibfnamefont {J.~G.}\
			\bibnamefont {Analytis}},\ }\bibfield  {title} {\bibinfo {title} {Exchange
			biased anomalous hall effect driven by frustration in a magnetic kagome
			lattice},\ }\href {https://doi.org/10.1038/s41467-020-14326-9} {\bibfield
		{journal} {\bibinfo  {journal} {Nature Communications}\ }\textbf {\bibinfo
			{volume} {11}},\ \bibinfo {pages} {560} (\bibinfo {year} {2020})}\BibitemShut
	{NoStop}%
	\bibitem [{\citenamefont {Menil}\ \emph {et~al.}(2024)\citenamefont {Menil},
		\citenamefont {Leridon}, \citenamefont {Cavanna}, \citenamefont {Gennser},
		\citenamefont {Mailly}, \citenamefont {Ding}, \citenamefont {Li},
		\citenamefont {Zhu}, \citenamefont {Fauqu\'{e}},\ and\ \citenamefont
		{Behnia}}]{MenilCondMat24}%
	\BibitemOpen
	\bibfield  {author} {\bibinfo {author} {\bibfnamefont {C.}~\bibnamefont
			{Menil}}, \bibinfo {author} {\bibfnamefont {B.}~\bibnamefont {Leridon}},
		\bibinfo {author} {\bibfnamefont {A.}~\bibnamefont {Cavanna}}, \bibinfo
		{author} {\bibfnamefont {U.}~\bibnamefont {Gennser}}, \bibinfo {author}
		{\bibfnamefont {D.}~\bibnamefont {Mailly}}, \bibinfo {author} {\bibfnamefont
			{L.}~\bibnamefont {Ding}}, \bibinfo {author} {\bibfnamefont {X.}~\bibnamefont
			{Li}}, \bibinfo {author} {\bibfnamefont {Z.}~\bibnamefont {Zhu}}, \bibinfo
		{author} {\bibfnamefont {B.}~\bibnamefont {Fauqu\'{e}}},\ and\ \bibinfo
		{author} {\bibfnamefont {K.}~\bibnamefont {Behnia}},\ }\href
	{https://arxiv.org/abs/2407.11836} {\bibinfo {title} {Magnetic memory and
			distinct spin populations in ferromagnetic co$_3$sn$_2$s$_2$}} (\bibinfo
	{year} {2024}),\ \Eprint {https://arxiv.org/abs/2407.11836} {arXiv:2407.11836
		[cond-mat.mtrl-sci]} \BibitemShut {NoStop}%
	\bibitem [{\citenamefont {\v{Z}ivkovi\'{c}}\ \emph {et~al.}(2023)\citenamefont
		{\v{Z}ivkovi\'{c}}, \citenamefont {Kassem}, \citenamefont {Tabata},
		\citenamefont {Waki},\ and\ \citenamefont {Nakamura}}]{ZivkovicJPSJ23}%
	\BibitemOpen
	\bibfield  {author} {\bibinfo {author} {\bibfnamefont {I.}~\bibnamefont
			{\v{Z}ivkovi\'{c}}}, \bibinfo {author} {\bibfnamefont {M.~A.}\ \bibnamefont
			{Kassem}}, \bibinfo {author} {\bibfnamefont {Y.}~\bibnamefont {Tabata}},
		\bibinfo {author} {\bibfnamefont {T.}~\bibnamefont {Waki}},\ and\ \bibinfo
		{author} {\bibfnamefont {H.}~\bibnamefont {Nakamura}},\ }\bibfield  {title}
	{\bibinfo {title} {Doping dependence of the in-plane transition in
			co3sn2s2},\ }\href {https://doi.org/10.7566/JPSJ.92.095001} {\bibfield
		{journal} {\bibinfo  {journal} {Journal of the Physical Society of Japan}\
		}\textbf {\bibinfo {volume} {92}},\ \bibinfo {pages} {095001} (\bibinfo
		{year} {2023})},\ \Eprint
	{https://arxiv.org/abs/https://doi.org/10.7566/JPSJ.92.095001}
	{https://doi.org/10.7566/JPSJ.92.095001} \BibitemShut {NoStop}%
	\bibitem [{\citenamefont {Clark}\ \emph {et~al.}(1995)\citenamefont {Clark},
		\citenamefont {Hanson}, \citenamefont {Lefloch},\ and\ \citenamefont
		{Segransan}}]{Clark}%
	\BibitemOpen
	\bibfield  {author} {\bibinfo {author} {\bibfnamefont {W.~G.}\ \bibnamefont
			{Clark}}, \bibinfo {author} {\bibfnamefont {M.~E.}\ \bibnamefont {Hanson}},
		\bibinfo {author} {\bibfnamefont {F.}~\bibnamefont {Lefloch}},\ and\ \bibinfo
		{author} {\bibfnamefont {P.}~\bibnamefont {Segransan}},\ }\bibfield  {title}
	{\bibinfo {title} {Magnetic resonance spectral reconstruction using
			frequency-shifted and summed fourier transform processing},\ }\href@noop {}
	{\bibfield  {journal} {\bibinfo  {journal} {Rev. Sci. Instrum.}\ }\textbf
		{\bibinfo {volume} {66}},\ \bibinfo {pages} {2453} (\bibinfo {year}
		{1995})}\BibitemShut {NoStop}%
	\bibitem [{\citenamefont {Bussandri}\ and\ \citenamefont
		{Zuriaga}(1998)}]{Bussandri}%
	\BibitemOpen
	\bibfield  {author} {\bibinfo {author} {\bibfnamefont {A.~P.}\ \bibnamefont
			{Bussandri}}\ and\ \bibinfo {author} {\bibfnamefont {M.~J.}\ \bibnamefont
			{Zuriaga}},\ }\bibfield  {title} {\bibinfo {title} {Spin-echo mapping
			spectroscopy applied to nqr},\ }\href@noop {} {\bibfield  {journal} {\bibinfo
			{journal} {Journal of Magnetic Resonance}\ }\textbf {\bibinfo {volume}
			{131}},\ \bibinfo {pages} {224 } (\bibinfo {year} {1998})}\BibitemShut
	{NoStop}%
\end{thebibliography}

\providecommand{\noopsort}[1]{}\providecommand{\singleletter}[1]{#1}%

\newpage
\begin{center}\textbf{Supplementary information}\end{center} 

\subsection{Samples}

Single crystals of magnetic Weyl semimetal Co$_3$Sn$_2$S$_2$ were grown by the self-flux method Ref.~\onlinecite{LohaniPRB23}. For the NMR measurements at T=300~K the oriented single crystal with m=17.6~mg was used. For the ZFNMR measurements at low temperatures for better penetration of the high-frequency magnetic field the set of small crystals with m$\approx$60~mg was crushed in an agate mortar into the powder with a particle size of the order of 50~$\mu m$ and packed in Stycast 1266A epoxy resin.

\subsection{NMR technique}

The NMR measurements were done using a home-built coherent pulsed NMR
spectrometer. NMR or ZFNMR spectra were taken "point by point"
with a $\pi /2-\tau -\pi$ radio frequency (RF) pulse
sequence by varying the magnetic field or frequency, correspondingly, in equal steps. For the spectra shown in Fig.~2(a-c) of the main text, we used a fixed frequency of 67.2~MHz and a sweep field range 6.1~T to 7.5~T. In both modes the full spectra were then constructed using a Fourier mapping algorithm~\cite{Clark,Bussandri}. The usual $\pi /2$ pulse length was 0.8~$\mu s$. The amplitude of the RF pulses were adjusted for the maximum of the observed spin-echo signal at the central part of the spectrum. The minimum practical $\tau $ values used in our experiments was 10~$\mu s$. For the ZF-NMR measurements sample was cooled in the zero field.

Generally a NMR spectrum in a ferromagnetic system can be composed of a mixture of signals coming from domains and domain walls, and the their contribution to the final amplitude of the spin echo will depend on the relative volumes of domain and domain walls, the local anisotropy, the exchange energy, and many others \cite{TurovPetrov}. The fact that we observed a rather narrow and single  $^{59}$Co ZFNMR line at all temperatures indicates that this is a signal from domains.

In magnetically ordered state due to the nuclear-electronic hyperfine coupling the applied RF field induces oscillations of the electronic moments. These oscillations lead to enhancement of the RF field at the nuclear site and of the back response of nuclear magnetization - spin-echo signal. The enhancement factor $\eta$ is different in domain and domain wall. For domains $\eta_D$ can be expressed by the relation $\eta_D = B_{int}/(B_{a} + B_{0})$, where $B_{int}$, $B_{a}$ and $B_{0}$, are the internal, the anisotropy, and the applied fields, respectively, and usually ranges from 10 to 100 \cite{TurovPetrov}. In our experiments we found that for $^{59}$Co ZFNMR in the Co$_3$Sn$_2$S$_2$ compound the enhancement factor $\eta$ is equal to one as the maxima of the spin-echo signal in paramagnetic and ferromagnetic states were observed at the same amplitude of the RF pulses. This clearly indicates a large value of the anisotropy field $B_{a} \sim B_{int}$ in Co$_3$Sn$_2$S$_2$.

\subsection{Fitting procedure}

Here we describe the procedure used to obtain the fits shown in Fig.~2(a-c) of the main text in order to extract the magnetic $\hat{K}$ and quadrupolar $\hat{V}$ tensors. The principal values of both tensors represent 5 parameters - $K_x, K_y, K_z, \nu_Q$ and $\eta$ - as the three components of EFG tensor are linked by Laplace equation. The orientation of the principal axes of the $\hat{K}$ and $\hat{V}$ tensors was defined using two sets of Euler angles $(\phi_{K_0},\theta_{K_0},\psi_{K_0})$ and $(\phi_{V_0},\theta_{V_0},\psi_{V_0})$, correspondingly, and $ZYZ$ convention for their rotation. In order to maintain mirror plane symmetry of the Co site in the kagome sublattce the orientations of the the $K_y$ and $V_{YY}$ axis were fixed perpendicular to the mirror plane, which corresponds to the fixed angles values: $\phi_{K_0} = \phi_{V_0} = 180^{\circ}$ and $\psi_{K_0} = \psi_{V_0} = 90^{\circ}$. Therefore only angles $\theta_{K_0}$ and $\theta_{V_0}$ were adjustable parameters. To describe the two other unequivalent Co sites of the kagome sublattice the angles $\psi_{K_0}$ and $\psi_{V_0}$ were changed by 120$^{\circ}$ and 240$^{\circ}$, correspondingly.

The procedure consists of simultaneous fitting of all line positions of the whole set of 12 spectra obtained at 300~K for a single crystal at different orientations of the applied magnetic field. The direction of external magnetic field $\vec{B}_{0}$ is defined by the spherical angular coordinates $\theta$ and $\varphi$ in respect to crystallographic axis $a$ and $c$ - see the insert in fig.~2(a). Since there are three Co sites, for each such set the calculation involves 12x3 diagonalizations of the hamiltonian Eq.~1 in the main text. In principle the orientation of the magnetic field is well known for each spectrum (the crystal was oriented by x-rays), however, since the line positions are extremely sensitive to this orientation, we found that in order to obtain convergence on the whole set of spectra, it was necessary to add a few more adjustable parameters to take into account small misalignment errors in positioning of the crystal on the sample holder as well as the direction of the rotation axis. In total we have 10 parameters to fit the whole set of spectra.

Note that direct fitting of the spectral shapes would be impractical or even impossible because, as already mentioned in  the main text, the experimental setup does not allow an accurate determination of the relative line intensities of different transitions. Therefore dataset used for fitting was a list of sets of frequencies observed at different sample orientations.  This approach adds a little challenge to properly define a fitness function ($\chi^2$), since for many spectra the number of recorded peaks is less than the theoretical number of 21 because of peak overlapping. After some trials we came up with a formula for $\chi^2$ inspired from the so-called Hausdorff distance (sort of measure of "distance" between sets of different cardinalities).

The difficulty of the fitting problem is not only due to the number of parameters and the amount of calculations needed for simulations, probably the main difficulty comes from the necessarily "noisy" character of the $\chi^2$ function, wiggling rapidly as peaks move when modifying any of the fit parameters.
Obviously, gradient-based optimization approaches do not work well in such cases, on the other hand the Differential Evolution (DE) method has been shown to be very effective in \textit{global} optimization problems and is therefore the method of choice here (for a review see references in the main text). Even though the DE method is computationally intensive, one can take advantage of the fact that it is
parallel in nature since a population of trial vectors can be tested by independent threads. Therefore it is perfectly adapted for computations on GPU, yielding a significant speedup compared to single CPU. Here the model and $\chi^2$ calculations were coded in OpenCL and run on GPU. We have used the NVIDIA RTX 3060 GPU (a mid-range gaming GPU featuring about 3500 cores),  allowing to perform several hundred thousand diagonalizations per second. The typical population size was of order of $10^5$, the convergence often needed a few thousand iterations (i.e. testing a total of $10^8$ trial vectors or so). We have run the calculations many times for different initial parameters and their ranges to check for multiple possible solutions.  To double-check  the correctness of the calculations, the obtained spectra were compared against those calculated with a different,  independently developed Python-based, simulation software.

\subsection{Hyperfine Field}

We want to detail here how the magnetic internal field \Bint~is related to the electronic susceptibility or magnetization. In the paramagnetic phase, we have determined the magnetic shift tensor $\hat{K}$ at 300~K. Its strongest component $K_z$ is found almost along $c$, while the value $K_y$ is twice smaller and $K_x$ nearly zero. As shown in Fig. S1, the susceptibility is actually larger in-plane than along $c$ at 300~K, suggesting a stronger coupling to the out-of plane value. As the difference between $K_x$ and $K_y$ is averaged over the 3 Co sites in the susceptibility measurement, it is not possible to resolve the in-plane couplings. Moreover, there are generally 2 contributions to the susceptibility, from orbital and spin parts. The components of the magnetic shift tensor are linked to the macroscopic susceptibility $\chi$ through specific hyperfine couplings $A$
\begin{equation}
	\hat{K}=\hat{K}_{s}+\hat{K}_{orb}=\hat{A}_{s}\hat{\chi}_{s}(T)\,+\,\hat{A}_{orb}\hat{\chi}_{orb}.
\end{equation}
It is very likely that the orbital part dominates the susceptibility at 300~K, while the spin part certainly dominates it in the magnetic phase, explaining the opposite anisotropies. Therefore, the value of $\hat{K}$ measured at high temperatures may not be relevant for the analysis of the low temperature magnetic phase. 

In the magnetic state, with no applied field, we simply have :

\begin{equation}
	\vec{B}_{int}=\hat{A}_{s}\vec{M}_{s}
	\label{eq2}
\end{equation}%
where $M_{s}$ is the electronic magnetic moment. If $\hat{A}_{s}$ is isotropic, all we say about \Bint~can be directly transposed to the magnetic moment. However, if there is a large anisotropy, for example a larger value along $c$ $A_z>A_x$, as for the shift tensor we determined at 300~K, we would underestimate the in-plane component. Indeed, assuming that $\hat{A}$ and $\hat{K}$ share the same orientation and neglecting the small tilt of these tensors with respect to the $c$ axis, the relation between the tilt angles $\theta$ of $\vec{B}_{int}$ and $\theta_M$ of the Co moment, reads $$\tan \theta = \frac{A_x}{A_z} \tan \theta_M.$$ This is why the value of the tilt of $\vec{B}_{int}$ we propose below $T_c$ and $90$~K could be lower bounds for the corresponding values of $\vec{M}_{s}$. 

\begin{figure}[h]
	\center
	\includegraphics[width=0.9\linewidth]{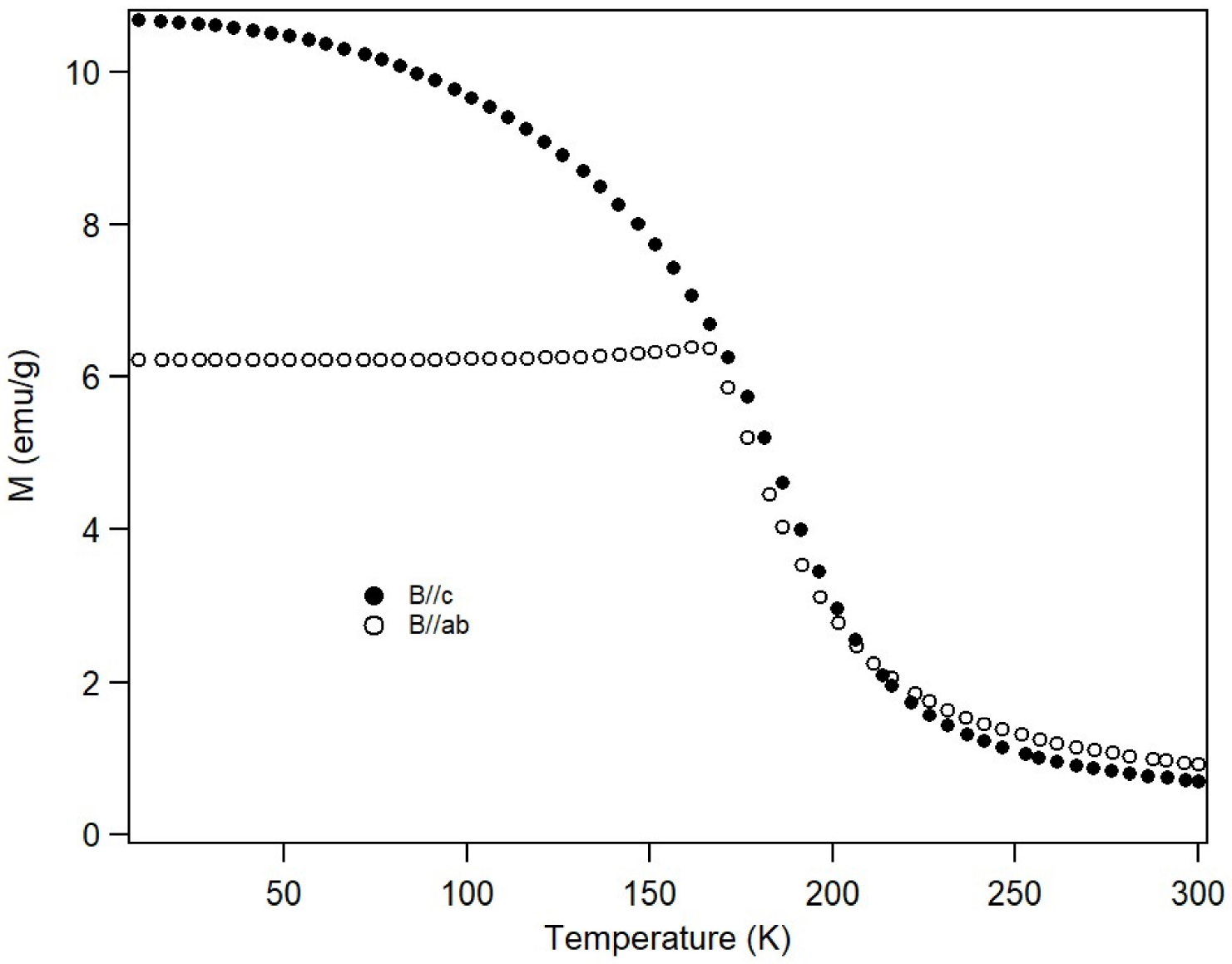}
	\caption{SQUID magnetization measured at 7~T in a single crystal in two perpendicular orientations (field cooled).}
	\label{fig:FigS1}
\end{figure}

\end{document}